\documentclass[a4paper,10pt]{article}
\usepackage[nointlimits]{amsmath}
\usepackage{amsfonts}
\usepackage{amssymb}
\usepackage{amsmath}
\usepackage{amsthm}
\usepackage{latexsym}
\usepackage{graphicx}
\usepackage{layout}
\usepackage[english]{babel}
\usepackage{fancyhdr}

\usepackage{lscape} 

\usepackage{amscd}

\usepackage{color}

\newtheorem{lemma}     {Lemma}[section]
\newtheorem{thm}   [lemma]{Theorem}
\newtheorem{teorema1}   [lemma]{Theorem}
\newtheorem{prop}      [lemma]{Proposition}
\newtheorem{coro}    [lemma]{Corollary}
\newtheorem{cong1}      [lemma]{Conjecture}
\newtheorem{remark1}    [lemma]{Remark}

\numberwithin{equation}{section}

\newcommand{\sgn}{\text{sign}}

\renewcommand{\(}{\left(}        \renewcommand{\)}{\right)}
\renewcommand{\[}{\left[}        \renewcommand{\]}{\right]}

 \renewcommand{\d}{\delta}
 
 \renewcommand{\a}{\alpha}
 \renewcommand{\b}{\beta}
 \newcommand{\g}{\gamma}
  \renewcommand{\o}{\omega}
\renewcommand{\t}{\theta}

\newcommand{\dis}{\displaystyle}

\newcommand{\mmmintone}[1]{{\dis{\int\kern -.43cm
-}}_{\kern-.21cm\substack{#1}}\;\;}
\newcommand{\mmmintwo}[2]{{\dis{\int\kern -.43cm
-}}_{\kern-.21cm\substack{#1}}^{\substack{#2}}\;\;}
\newcommand{\submint}{{\scriptstyle{\int\kern -.66em -}}}
\newcommand{\submintone}[1]{{\scriptstyle{\int\kern -.66em
-}}_{\scriptscriptstyle{\kern-.21em\substack{#1}}}}
\newcommand{\fracmint}{{\textstyle{\int\kern -.88em -}}}
\newcommand{\fracmintone}[1]{{\textstyle{\int\kern -.88em
-}}_{\scriptscriptstyle{\kern-.21em\substack{#1}}}\;}

\newcommand{\ra}{\rightarrow}

\renewcommand{\P}{\mathbf{P}}
\newcommand{\E}{\mathbf{E}}
\renewcommand{\v}{\mathbf{var}}

\newcommand{\eps}{\epsilon}

\newcommand{\ga}{\gamma}

\newcommand{\la}{\lambda}

\renewcommand{\i}{\infty}

\newcommand{\nada}[1]{}

\newcommand{\lap}{\la_{_+}}
\newcommand{\lam}{\la_{_-}}

\newcommand{\lapm}{\la_{_\pm}}

\newcommand{\Pe}{\mathbf{P}}

\title{Langevin dynamics with a tilted periodic potential}

\author{
\emph{Gioia Carinci$^{\textup{{\tiny(a)}}}$ \hskip.5cm and \hskip.5cm   
Stephan Luckhaus}$^{\textup{{\tiny(b)}}}$\\
\\
{\emph{\small $^{\textup{(a)}}$
University of Modena and Reggio Emilia, \small via G. Campi 213/b, 41125 Modena, Italy}}\\
{\emph{ \small{e-mail: gioia.carinci@unimore.it}}}
\\ \\
{\emph{\small $^{\textup{(b)}}$ University of Leipzig, Augustus Platz, 10-11, D-04109 Leipzig, Germany }}\\
{\emph{ \small{e-mail: luckhaus@mis.mpg.de}}}
}

\date{ }

\begin{document}

\maketitle

\begin{abstract}
We study a Langevin equation for a particle moving in a periodic potential
    in the presence of viscosity $\ga$ and subject to a further external field $\a$. 
    For a suitable  choice of the parameters $\a$ and $\ga$ the related deterministic dynamics  yields heteroclinic  orbits. In such a  regime, in absence of
    stochastic noise   both confined and unbounded orbits coexist.
    We prove that, with the inclusion of an arbitrarly small  noise only the
    confined orbits survive in a sub-exponential time scale.
 \end{abstract}





\section{Introduction}

In many physical contexts e.g. the switching of magnetic domain walls (Barkhausen noise) or the motion of twin boundaries in crystals, one observes an intermittent dynamics of energy relaxation with ``relaxation events'' of random amplitude. This type of dynamics  is commonly called Avalanche Dynamics (see e.g. \cite{PM,PM1}).
In order to  give a possible explanation for this kind of mechanism, we need three ingredients:
a rough interaction potential with many local energy barriers, and a small tilt;
an  ``almost Hamiltonian'' dynamics that approximately conserves the total energy, helping the system to overcome the next  energy barrier once it passes the first one; finally a weak coupling to a heat bath. 
The simplest model  one may consider  in a two dimensional phase space taking into account all these factors is the Brownian motion  in a tilted periodic potential.
 Let $x(t) \in \mathbb{R}$ be  the coordinate of a one-dimensional Brownian particle moving in
a periodic oscillating  potential $V_0(x)$ (e. g. $V_0(x)=\cos x$) in the presence of viscosity $\ga$ and  subject to  an
additional constant external force $\a$. The dynamics for $x(t)$ is
governed by the Langevin equation:
\begin{equation}\label{eq:0}
\ddot{x}+\gamma \, \dot{x}+V_0'(x)= \a+ \eps\dot{w}(t)
\end{equation}
where $\dot{w}(t)$ is  the
white noise, and $\eps$  a small parameter.
The  model is not  new,  indeed 
it has been widely studied  because of its well known applications 
 to  several classical  areas of physics such as electronics (see e.g. \cite{StrI, StrII})
and solid state physics (superionic conductors and Josephson tunneling junction, 
see e.g. Section 11.1 of \cite{Ri}). 


According to the values of the parameters  and the initial conditions, 
the particle may escape in the direction of the   force $\a$ 
or be trapped for a long time in one of the wells  of the potential. 
 Without any noise ($\eps=0$), when $\a$ is large enough there are only ``running solutions'', i.e. unbounded solutions. 
When the force $\a$ is small enough and the friction parameter $\ga$ is large enough, the
particle finally reaches one of the minima of the potential. In this case
there are only ``locked solutions''. For $\a$ and $\ga$ small enough, both types of solution coexist.
With the addition of the noise there are certainly  transitions between the locked and the running
state, due  to  large deviations effects and related to the problem of diffusion exit from a domain \cite{FW}. These kind of transitions  take place in the Kramer's escape time scale, i.e. the time required for a brownian particle to escape from a well in the presence of viscosity,  that is of order of $\exp(\gamma /\eps^2)$ for $\eps^2<<\gamma<<1$ (see \cite{Kr}). 
There is a wide literature concerning  the diffusion along a periodic potential, mostly inspired by the Kramer's theory \cite{Kr},  
 that includes both numerical and theoretical studies (see e.g. \cite{HP,PV,Ri,RV}).
 In the present paper we do not assume a  large deviations point of view, since we are interested in a different time scale. 
 In certain critical regimes, due to the instability of the dynamics, small stochastic perturbations may indeed affect the macroscopic behavior even in a faster time scale.
This for instance turns out to be the case for the particular choice $\a=\a_\gamma$ yielding heteroclinic deterministic orbits (i.e. orbits  connecting two consecutive maxima of the potential). 
In the present work we show that, by stochastically perturbing the system in the critical regime $\a=\a_\ga$ that exibits both running and locked solutions, 
only the locked solutions survive in a subexponential time scale (and more precisely if one observes the system for times $\ln{\eps^{-1}}<<T<<\exp(\gamma /\eps^2)$).
The introduction of an arbitrarly small noise has thus a macroscopic effect, since it reduces the bistability region even in this ``fast time scale''.
The next case to consider is thus the region $\a>\a_\ga$. In particular one may wonder whether there exists a transition regime from the fast to the slow time scale where only the Kramer's diffusion is observed.


\vskip.5cm





\subsection*{Deterministic orbits}

We denote by $V(x)$ the total potential    taking into account also the linear term due to the
external force,  $V(x)=V_0(x)-\a x$.
The equation of the motion \eqref{eq:0} then
becomes
\begin{equation}\label{eq:1}
\ddot{x}+\gamma \, \dot{x}+V'(x)= \epsilon \, \dot{w}(t)
\end{equation}
and the related first order system is 
\begin{equation}\label{eq:x,p}
\left\{
\begin{array}{ll}
  \dot{x}  = p \\
  \dot{p}  =-\gamma p - V'(x)+ \epsilon \dot{w}.
\end{array}
\right.
\end{equation}

 Consider the
deterministic system related to \eqref{eq:x,p}, i.e. the $\epsilon=0$ case,
\begin{equation}\label{eq:X,P}
\left\{
\begin{array}{lll}
  \dot{X}  = P \\
  \dot{P}  =-\gamma P - V'(X)
\end{array}
\right.
\end{equation}
and  sketch the phase diagram. In each period of the original potential $V_0(x)$ there are confined and escaping orbits. 
The picture changes  according to $\a$ and
$\gamma$.

\begin{figure}[!htbp]
\mbox{%
\begin{minipage}{.40\textwidth}
\centering
\includegraphics[width=73mm, height=35mm]{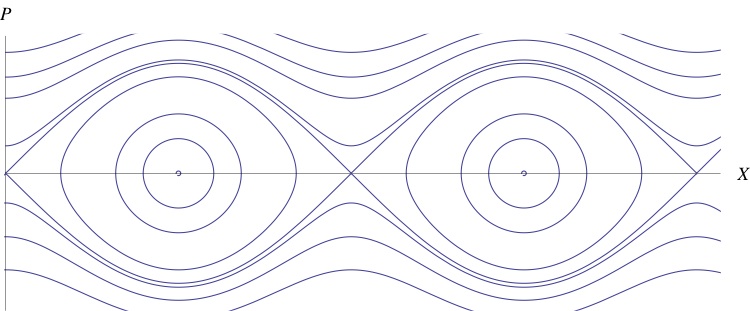}
\caption{Phase diagram for $\a=\ga=0$. The orbits are periodic.}\label{Periodico}
\end{minipage}%
\hskip2cm
\begin{minipage}{.40\textwidth}
\centering
\includegraphics[width=73mm, height=35mm]{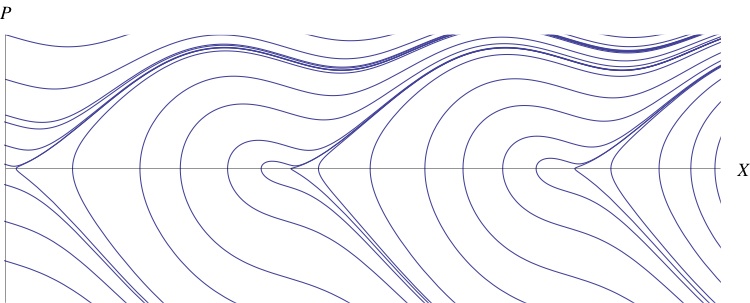}
\caption{Phase diagram for  $\a>1$. There are only running orbits.}\label{Phase2}
\end{minipage}%
}
\end{figure}

In absence of viscosity and external force (i.e. for $\ga=\a=0$),
the orbits are periodic (see Figure \ref{Periodico}).

For positive $\gamma$ and $\a$ the dynamics loses its periodicity.
 The inclusion of the friction makes the system dissipative,
the particles lose energy, and the confined orbits are attracted
by the local minima of the potential, whereas the running solutions have an asymptotic
effective velocity.

\begin{figure}[!htbp]
\mbox{%
\begin{minipage}{.41\textwidth}
\centering
\includegraphics[width=73mm, height=35mm]{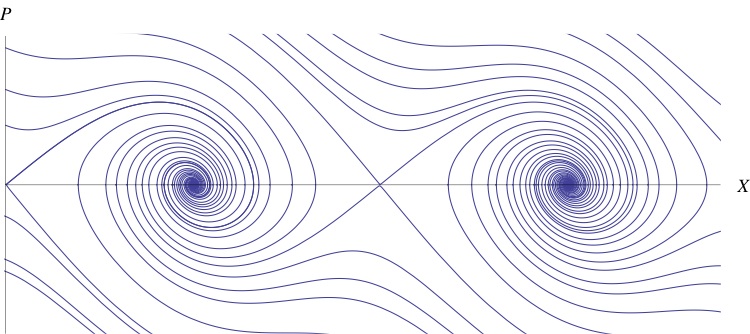}
\caption{Phase diagram for $\ga>0$ and  $0<\a< \a_\ga$. There are only confined orbits.}\label{Phase}
\end{minipage}%
\hskip2cm
\begin{minipage}{.41\textwidth}
\centering
\includegraphics[width=73mm, height=35mm]{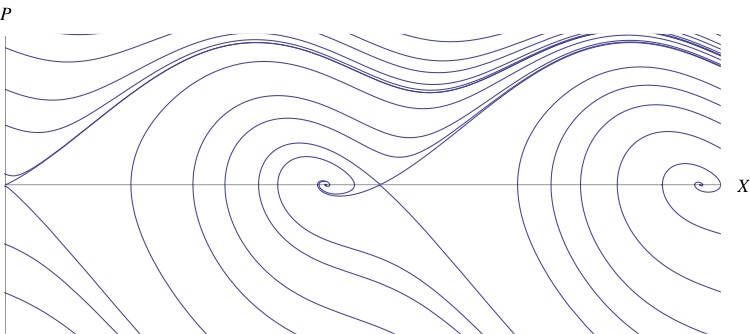}
\caption{Phase diagram for $\ga>0$ and  $\a_\ga<\a <1$. There are both running and and confined orbits.}\label{Phase1}
\end{minipage}
}
\end{figure}

Figures \ref{Phase2}, \ref{Phase} and \ref{Phase1} provide three possible phase diagrams.
For $\a>1$ the total potential $V(x)$ does not have local minima and then there are only   running solutions (see Figure \ref{Phase2}).

When $\a< 1$, there are also bounded solutions.
For any $\ga>0$, there exists a critical value $\a_\ga>0$ such that, for $0<\a< \a_\gamma$,  there are only confined orbits (see Figure \ref{Phase}).
For $\a_\ga<\a< 1$ running and bounded orbits coexist (see Figure \ref{Phase1}).

The situation is well resumed in Figure \ref{Alpha} that shows the dependence on $\ga$ of the critical value $\a_\ga$. As we will see below,   $\a_\gamma$ is of the order of $\gamma$ as $\gamma \to 0$.

For any $\a <1$ there exist  critical orbits, i.e.
orbits asymptotically converging to some saddle equilibria  (corresponding to some local maxima of the potential) in the phase plane.
The critical scaling $\a=\a_\ga$  is the one that gives rise to heteroclinic orbits, i.e.
orbits  connecting two consecutive local maxima of the potential.

Each orbit  can be piecewise expressed
by a  function in the phase plane that we will usually denote by $\wp(x)$. 
We say that $\wp(x)$ is an orbit of our dynamics if there exists a solution $(X(t),P(t))$ of \eqref{eq:X,P}
  and a suitable time interval $I \subseteq (0,+\i)$  such that $\wp(X(t))=P(t)$ for any $t \in I$. 
 $\wp(x)$ must verify the 
\begin{equation}\label{eq:p}
 \frac {d\wp}{dx}\,(x)=-\gamma -\frac {V'(x)}{\wp(x)}
\end{equation}
 

\begin{figure}[!htbp]
\mbox{%
\begin{minipage}{.40\textwidth}
\includegraphics[width=70mm]{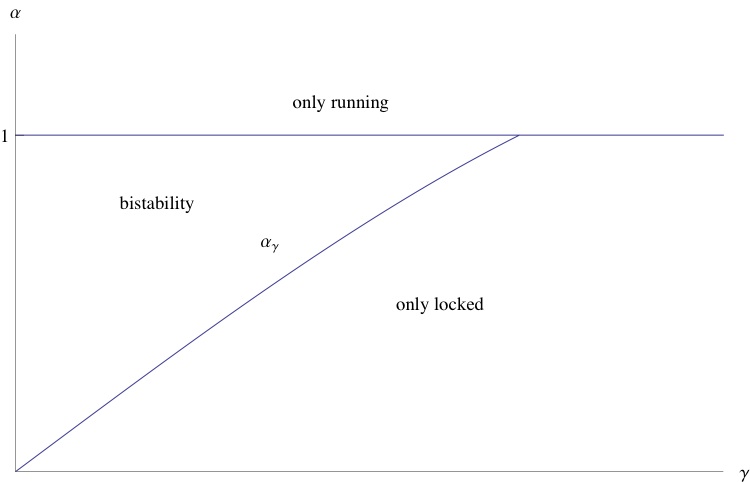}
\caption{The picture shows the three different regimes in the plane $(\ga, \a)$.}\label{Alpha}
\end{minipage}%
\hskip2cm
\begin{minipage}{.40\textwidth}
\includegraphics[width=70mm]{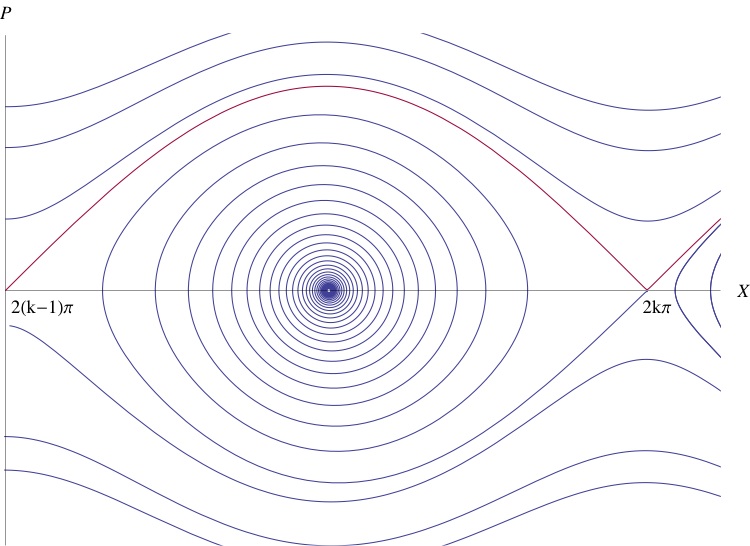}
\caption{When $\a=\a_\ga$ the heteroclinic orbit 
$\wp_k^*(x)$   (red line) separates the escaping
 and the confined solutions.}\label{Hete2}
\end{minipage}%
}
\end{figure}


\subsection*{The problem}

In the present paper we are interested  in the critical regime ($\a=\a_\ga$) dynamics  in the 
small noise limit ($\eps \to 0$).
Thus  from now on we fix $\ga$ small enough and $\a=\a_\ga$. We denote by $\wp^*_k(x)$ the $k$-th heteroclinic orbit,
i.e. the orbit connecting the $k-1$-th maximum of the potential with the $k$-th one:
\begin{equation}\label{eq:pBC}
  \lim_{x \to 2(k-1)\pi^+}\wp_k^*(x)=\lim_{x \to 2 k\pi^-}\wp_k^*(x)=0
\end{equation}
 see Figure \ref{Hete2}.

In this paper we are not concerned with large deviations. We investigate the problem in a sub-exponential time scale. 
Far from the critical orbits we expect that the noise does not macroscopically affect  the deterministic dynamics in such a fast time scale.
On the other hand, there may be a macroscopic perturbation of the deterministic dynamics in a neighbourhood of the heteroclinic orbits.
Then we choose the initial value to lay  on
one of the heteroclinic orbits, i.e. we denote by  $(x(t),p(t))$ the solution of the problem
\begin{equation}\label{eq:x,p'}
\left\{
\begin{array}{ll}
  \dot{x}  = p  & x(0)=-\pi\\
  \dot{p}  =-\gamma p - V'(x)+ \epsilon \dot{w} &
  p(0)=\wp^*_0(-\pi),
\end{array}
\right.
\end{equation}
 and study the probability law of $(x(t),p(t))$ in the
limit as $\eps \to 0$.

We show that, at each time, the probability for the particle, to
get across the next well is $1/2$, in the limit as $\eps \to 0$.
We prove, thus,  that for fixed $\ga$ small enough, the random
variable associated to the number of wells passed by $(x(t),
p(t))$ has, for vanishing $\epsilon$, a geometric distribution of
parameter $1/2$. This implies, in particular, that the particle
will finally be trapped in one of the wells for a long time, with
 probability 1 as $\eps \to 0$. The involved time scale, or more precisely, the velocity the 
 particle travels before to be trapped in one of the wells is of order $\ln \eps$.
With the inclusion of the noise, thus, the bistability between the
locked and the running states is lost and only the locked state
survives in this ``fast time scale''.

\subsection*{Notation}
Before turning to the precise statement, we introduce some notations. We shall use

\begin{itemize}
\item $f(\d)=o(g(\d))$ as $\d \to 0$ if $\lim_{\d \to 0} f(\d)/g(\d)=0$; 

\item $f(\d)=\mathcal O(g(\d))$ as $\d \to 0$  if  there exists $c>0$ such that $|f(\d)|\le c |g(\d)|$ for any sufficiently small $\d$;

\item $f(\d)= \Theta(g(\d))$ as $\d \to 0$  if there exist $c_1,c_2>0$ such that $c_1|g(\d)|\le |f(\d)|\le c_2 |g(\d)|$ for any sufficiently small $\d$;

\item $X \sim \mu$ if the stochastic variable has probability law $\mu$.

\end{itemize}

\subsection*{Basic parameters}

Although our results do not depend on
   the exact form of the potential,
   for the sake of simplicity we choose $$V(x)=\cos (x-x_\a)-\a(x-x_\a), \quad \quad \quad x_\a= \arcsin \a$$ 
 in such a way that $V(x)$ attains its local maxima at $x=2k\pi$ and its local minima at $x=(2k-1)\pi +2x_\a$.

Let us investigate how $\a_\g$ is related to $\g$.  All the orbits
$\wp_k^*(x)$ must verify the equation \eqref{eq:p}, then
\begin{equation}\label{eq:p^2}
\frac 1 2 \, \frac{d}{dx}\,(\wp^*_k(x))^2=-V'(x)-\gamma
\wp_k^*(x).
\end{equation}

\vskip.3cm

By integrating \eqref{eq:p^2} in $(2(k-1)\pi,2k\pi)$, and using
the asympthotic conditions \eqref{eq:pBC} we get
\begin{equation*}
2 \pi \, \frac{\a_\g}{\g}=\int_{2(k-1)\pi}^{2k\pi}\wp^*_k(x)\, dx=
 \Theta(1) \quad \quad \text{as} \quad\quad \ga \to 0
\end{equation*}
thus $\a_\g = \Theta(\g)$ in the limit as $\g \to 0$.

\vskip.3cm

We compute the asymptotic values of $d\wp^*_k(x)/dx$ for $x \to
2k\pi^\pm$. Let 
\begin{equation}\label{be}
\beta=\beta_\gamma:=-V''(2k\pi)=\sqrt{1-\a_\g^2}>0\quad \quad k
\in \mathbb{Z},
\end{equation}
then, from \eqref{eq:p} we obtain that the asympthotic slopes of the heteroclinic orbits:
\begin{equation}\label{la1}
\lap:=\frac d {dx}\,\wp_k^*(2(k-1)\pi^+) \quad\quad \text{and}\quad \quad
\lam:=\frac d {dx}\wp_k^*(2k\pi^-)
\end{equation}
must satisfy the equation
\begin{equation*}
\la^2+\g \la-\b=0,
\end{equation*}
then
\begin{equation}\label{la}
\lapm=\lapm^\g=\frac{-\g \pm \sqrt{\g^2+4\b_\g}}2
\end{equation}
notice that  $\lim_{\g \to 0}\b_\g=1$ and $\lim_{\g\to
0}\la_\ga^\pm=\pm 1$.
We define, moreover, the parameter
\begin{equation}\label{th}
\t=\t_\g:=\frac{|\lam|}{\lap}-1=\frac{2\g}{\sqrt{\g^2+4\b}-\g}
\end{equation}
 then $\t_\ga=\Theta(\g)$ as $\ga \to 0$ with $\lim_{\ga \to 0}\theta_\g/\g=1$.

\subsection*{Critical region dynamics}

In a neighborhood of the criticalities, i.e. when $x(t)$ approaches $2k\pi$, $V'(x(t))$ is well approximated by
$-\b(x(t)-2k\pi)$, since $V'(2k\pi)=0$. Hence the dynamics can be  approximated by the linear
system
\begin{equation}\label{eq:x,p'}
\left\{
\begin{array}{ll}
  \dot{x}  = p \\
  \dot{p}  =-\gamma p +\beta (x-2k\pi)+ \epsilon \dot{w}
\end{array}
\right.
\end{equation}

A convenient choice of  variables is given by
\begin{equation}\label{z,v}
\left\{
\begin{array}{ll}
  z_k(t):=p(t)-\lam(x(t)-2k\pi)\\
  v_k(t):=p(t)-\lap(x(t)-2k\pi)
\end{array}
\right.
\end{equation}

with $\lapm$ as defined in \eqref{la}. Since in these variables the linearized system \eqref{eq:x,p'} becomes
\begin{equation}\label{eq:z,v}
\left\{
\begin{array}{ll}
  \dot{z}  =  \lap z +\eps \dot{w}\\
  \dot{v}  =\lam v +\eps \dot{w}
\end{array}
\right.
\end{equation}
where the equations 
are coupled only per the stochastic term.

\vskip.3cm

Since $\wp_k^*(2k\pi)=0$,  by  \eqref{la1},
 for $|x(t)-2k\pi|$ small
enough we have
\begin{equation}\label{z:sim}
  z_k(t) = p(t)-\wp_k^*(x(t)) + \mathcal O((x(t)-2k\pi)^2)\quad \quad \text{for}\quad \quad  x(t)<2k\pi\\
\end{equation}
and
\begin{equation}\label{v:sim}
  v_k(t)=  p(t)-\wp_{k+1}^*(x(t)) +\mathcal O((x(t)-2k\pi)^2) \quad\quad \text{for}
  \quad \quad x(t)>2k\pi.
\end{equation}

The variable $|z_k(t)|$ can, thus, be thought of as a measure of the
distance, in the phase plane, from the $k$-th heteroclinic orbit
$\wp_k^*(\cdot)$ just before the $k$-th criticality. Equivalently $|v_k(t)|$ is a measure of
the distance  from the $k+1$-th heteroclinic orbit
$\wp_{k+1}^*(\cdot)$ just after the $k$-th criticality.

\subsection*{Stopping times}
Far from criticalities the dynamics is stable and
we expect the distribution of our process to be quite concentred in a
neighborhood of some deterministic paths (see the deterministic system \eqref{eq:X,P}).

One of the main technicalities  we run into in the proof of our result is the choice of the length of the critical interval. 
It is clear that the linear system \eqref{eq:x,p'} is a good approximation of our dynamics
as long as $x(t)$ is close enough to the $k$-th criticality $2k\pi$. This provides an upper bound 
on the critical interval length.
On the other hand, we need a good localization of the process at the beginning of the critical interval. We want the process to be still quite 
concetrated and not to diffuse too much. This clearly requires a lower bound on the length of the critical interval.

 We denote by $\eta_\eps$ the order of magnitude of the length of the critical interval.
 Then, in order to make valid our approximations we need to impose the following condition on $\eta_\eps$:
\begin{equation}\label{eq:nu}
\eta_\eps= \eps^\nu \quad \quad \quad \text{with} \quad \quad \quad \frac{(1+\t)^2}{3+2\t} <\nu <\frac {1}2
\end{equation}
and $\t$ as in \eqref{th}.

\vskip.2cm

We define two sequences of stopping times. We call $S_k$ the first time the process gets into the $k$-th
critical interval and by  $T_{k+1}$ the first exit time from the $k$-th
critical interval.  
The rigorous definition  is given as follows 
\begin{equation}\label{T,S}
\begin{array}{l}
S_k:=\inf \{ t \ge 0:\; v_k(t)\le \eta_\eps\}, \quad k \ge 0 \\
T_k:=\inf \{ t \ge S_{k-1}: \; |z_{k-1}(t)|\ge \eta_\eps\}, \quad
k \ge 1
\end{array}
\end{equation}
We will see that, under the condition \eqref{eq:nu},  both $|z_k(S_k)| $ and $|v_k(T_{k+1})|$ are $o(\eta_\eps)$ with large probability, then
$$
2k\pi-x(S_k)= \Theta(\eta_\eps) \quad \quad \text{and} \quad \quad  |x(T_{k+1})-2k\pi| =\Theta(\eta_\eps)
$$
with large probability, since $v_k(S_k)= |z_k(T_{k+1})|=\eta_\eps$ and
$x(\cdot)-2k\pi=[z_k(\cdot)-v_k(\cdot)]/(\lap-\lam)$.

\vskip.3cm

 At time
$S_k$ the fundamental variable is $z_k(S_k)$, since, as we showed before, $|z_k(S_k)|$
measures the distance from the $k$-th deterministic heteroclinic
path before the criticality.\\
At time $T_{k+1}$ the fundamental variable is $v_k(T_k)$ since $|v_k(T_{k+1})|$ is a measure of
the distance from the $k+1$-th heteroclinic path, or,  if the well has not been crossed,
a measure of the distance from  a suitable locked deterministic path.

$\text{sign}(z_k(T_{k+1}))$
establishes whether the solution has crossed the $k$-th  criticality or not. If
$z_k(T_{k+1})>0$ the $k$-th criticality has been passed by,
if $z_k(T_{k+1}) <0$  the
solution has been trapped in the $k-1$-th well.

\subsection*{The Result}

We investigate the probability law of the number of wells crossed
by $(x(t), p(t))$,  thus we define the random variable
\begin{equation*}
\mathcal N:= \inf \{k \ge 0 \, :\: z_k(T_{k+1})<0\}\in
\mathbb{N}\cup \{0\}
\end{equation*}
We look at the process stopped at time $T_{\mathcal N +1}$, $(x(t
\wedge T_{\mathcal N +1}),p(t \wedge T_{\mathcal N +1}))$. All the
processes labeled by $k$ defined in the previous Sections   (e.g.
$(z_k(t),v_k(t))$) are well defined for $k \le \mathcal N$. $S_k$
is defined for $k \le \mathcal N$ and $T_k$ for $k \le \mathcal
N+1$.

We denote by $\Pe$ the probability law of $(x(t \wedge
T_{\mathcal N +1}),p(t \wedge T_{\mathcal N +1}))$, 
the main result is given by the following Theorem.
\vskip.3cm

\begin{thm}\label{TEO}
There exists $c>0$ such that 
\begin{equation}\label{eq:TEO}
 \(\frac 12 - c \eps^\theta\)^{k+1}\le\Pe\left\{\mathcal N = k\right\}\le \(\frac 12 + c \eps^\theta\)^{k+1}
\end{equation}
for any  $k \in \mathbb{N} \cup \{0\}$ and $\eps$ small enough.
\end{thm}

\vskip.3cm
\begin{coro}
From Theorem \ref{TEO} it follows that  the r.v. $\mathcal N +1$ has, in the limit as $\eps \to 0$, a
geometric distribution of parameter $1/2$, i.e.
\begin{equation}\label{eq:TEO'}
\lim_{\epsilon \to 0}\Pe\left\{\mathcal N = k\right\}=\frac
1{2^{{k+1}}}
\end{equation}
This implies, in particular,  that the process crosses a finite number of wells:
\begin{equation}\label{eq:TEO''}
\lim_{\epsilon \to 0}\Pe\left\{\mathcal N  <\infty\right\}=1
\end{equation}

\end{coro}

\vskip.3cm

 $\mathcal N$ is the number of wells crossed by the
process, in the sense that the
first well  our process is trapped is the $\mathcal N$-th one,
i.e. the one where the potential has a local minimum at
$(2\mathcal N-1)\pi +2x_\a$.
 
\vskip.3cm

\begin{prop}\label{TEO0}
There exists $c>0$ such that
\begin{equation}\label{eq:TEO0}
\lim_{\epsilon \to 0}\Pe\left\{x(T_{\mathcal N +1})\le 2 \mathcal N \pi
-\frac{\eta_\eps}{(\lap-\lam)}\,(1-o(1)),\: \: p(T_{\mathcal N +1})<-\frac{\lap \eta_\eps}{(\lap-\lam)}(1-o(1)) \right\}= 1.
\end{equation}
\end{prop}

Once $(x(t),p(t))$ is trapped in the $\mathcal N$-well in the
sense of Proposition \ref{TEO0}, we expect that it remains
confined in it for a long time. It is quite reasonable to think that, due 
to the stochastic fluctuations, the process
will leave the well in a exponential time scale (order of $e^{\eps^{-1}}$).
This is not a problem we are concerned with, since, at this level,
we are looking at a faster   time scale, indeed, as we will see, the
 time required to get across a
criticality is of order $\ln (\eps^{-1})$. We say, thus, that
the process is ``confined in the $\mathcal N$-th well'' in the
sense that $x(t) \in (2(\mathcal N -1)\pi, 2 \mathcal N \pi)$,
after $T_{\mathcal N +1}$, for a time that is very long if
compared with $T_{\mathcal N +1}$ itself. For this reason it is
quite reasonable to  stop the process at time $T_{\mathcal N +1}$.

\vskip.3cm

\section{Definitions}
\label{sect:Intro}
The tecniques  used  to prove our result are based on a pathwise analysis of the stochastic trajectories,
(to this end see e. g. \cite{BG}).
We perform two different linearizations of the nonlinear equation \eqref{eq:x,p}. 
In the neighborhoods of the criticalities we use the linearization in the system \eqref{eq:x,p'} that is solved by a  gaussian process.
In the stable region $(x(t),p(t))$ is expected to track a suitable deterministic trajectory, thus 
linearization is done around a given deterministic solution.  Then we study the behavior of the approximating trajectories and use comparison techniques 
to get  from them a control on the original process.

\subsection*{Dynamics near the criticalities}

For $S_k$ as  in \eqref{T,S}, the process $(z_k(t),v_k(t))$
defined in \eqref{z,v} satisfies the problem
\begin{equation*}
\left\{
\begin{array}{ll}
  \dot z:=\lap z +\psi_k (x(t))+ \eps \dot w & \: z_k(S_k)=p(S_k)-\lam(x(S_k)-2k\pi)\\
  \dot v:=\lam v +\psi_k (x(t))+ \eps \dot w & \:
  v_k(S_k)=\eta_\eps
\end{array}
\right.
\end{equation*}

with
\begin{equation*}
\psi_k(x)=-V'(x)-\b (x-2k \pi)= 
\mathcal O((x-2k\pi)^2) \quad \quad \text{as} \quad x \to 2k\pi
\end{equation*}

We denote by $(\bar z_k(t), \bar v_k(t))$ the solution of the
related linear problem  \eqref{eq:z,v} starting from the same point: $(z_k(S_k), \eta_\eps)$.
Thus, for $t \ge S_k$,
\begin{equation}\label{zz}
\bar z_k(t)=z_k(S_k)\,e^{\lap (t-S_k)}+ \eps \,e^{\lap t}
\int_{S_k}^t e^{-\lap s} dw_s
\end{equation}
and
\begin{equation}\label{v}
 \bar v_k(t)=\eta_\eps\,
e^{\lam (t-S_k)}+ \eps \,e^{\lam t} \int_{S_k}^t e^{-\lam s}
dw_s.
\end{equation}

We define the errors
\begin{equation}\label{V,Z}
\mathcal V_k(t):=v_k(t)- \bar v_k(t) \quad\quad\quad \text{and} \quad\quad\quad
\mathcal{Z}_k(t):=z_k(t)- \bar z_k(t)
\end{equation}

thus
\begin{equation}\label{V}
\mathcal{V}_k(t)= e^{\lam t}\int_{S_k}^t e^{-\lam
s}\psi_k(x(s))\, ds \quad \quad\quad \text{and} \quad \quad\quad\mathcal{Z}_k(t)= e^{\lap t}\int_{S_k}^t e^{-\lap
s}\psi_k(x(s))\, ds
\end{equation}

\subsection*{Dynamics far from criticalities}

Suppose that $T_k<\i$ and that the solution has crossed the
$k-1$-th criticality, then we denote by $(X_k(t),P_k(t)), t \ge
T_k$ the deterministic path starting in $T_k$ from $(x(T_k),
p(T_k))$, i.e. the solution of the problem
\begin{equation}\label{X,P}
\left\{
\begin{array}{lll}
  \dot{X}  = P & X_k(T_k)=x(T_k)\\
  \dot{P}  =-\gamma P - V'(X) &  P_k(T_k)=p(T_k)
\end{array}
\right.
\end{equation}
then we expect that, as long as $x(t)$ is far enough from the criticalities, the dynamics 
is concentrated in a neighborhood of $(X_k(t),P_k(t))$.
We denote by
\begin{equation*}
y_k(t):=x(t)-X_k(t) \quad \quad\quad\quad \text{and} \quad\quad\quad\quad
q_k(t):=p(t)-P_k(t) 
\end{equation*}
thus $(y_k(t),q_k(t))$ is solution of 
\begin{equation}\label{eq:y,q}
\left\{
\begin{array}{ll}
  \dot{y}  =  q & y_k(T_k)=0\\
  \dot{q}  =-\g q-V''(X_t)y + \varphi(X_k(t),y)+\eps \dot{w} & q_k(T_k)=0
\end{array}
\right.
\end{equation}
 with
\begin{equation}\label{phi}
\varphi(X,y):=V''(X)y-[V'(X+y)-V'(X)]= \mathcal  O(y^2), \quad \quad \text{as} \quad |y|\to 0
\end{equation}
 
 \vskip.3cm
 
 Assuming, for the moment, that it is
possible, we define $\wp_k(x)$ as the curve on the phase plane
associated to $(X_k(t),P_k(t)), t \in [T_k,S_k]$, i.e. the
function such that $\wp_k(X_k(t))=P_k(t)$ for any $t \in
[T_k,S_k]$. We know from \eqref{eq:p}
 that $\wp_k(x)$ verifies the equation
\begin{equation}\label{eq:omega''}
\wp''(x)\wp(x)=-(\wp'(x))^2-\g\wp'(x)-V''(x),
\end{equation}
that is obtained by deriving \eqref{eq:p}. Let us define the function
\begin{equation}\label{o}
\o_k(t):= \frac d {dx}\,\wp_k(X_k(t))
\end{equation}
then, by using \eqref{eq:omega''}, we deduce that $\o_k(t)$  verifies the equation
\begin{equation}\label{eq:omega}
\dot{\o}=-\o^2-\g \o -V''(X_k(t)).
\end{equation}

 It turns out to be particularly convenient to pass to the variables
$(y_k(t),r_k(t))$, with
\begin{equation}\label{rr}
r_k(t):=q_k(t)-\o_k(t)y_k(t) \quad \quad t \ge T_k.
\end{equation}

thus $(y_k(t), r_k(t))$ is solution of the problem
\begin{equation}\label{eq:y,r}
\left\{
\begin{array}{ll}
  \dot{y}  =  r+\o_k(t) y & y_k(T_k)=0\\
  \dot{r}  =-(\g+\o_k(t))r +\varphi(X_k(t),y_k(t))+\eps \dot{w} & r_k(T_k)=0
\end{array}
\right.
\end{equation}

with $\varphi(X,y)$ as above. We denote by $(\bar y_k(t),\bar r_k(t))$ the solution of the associated
 linear problem:
\begin{equation}\label{eq:y,r}
\left\{
\begin{array}{ll}
  \dot{y}  =  r+\o_k(t) y & \bar y_k(T_k)=0\\
  \dot{r}  =-(\g+\o_k(t))r +\eps \dot{w} & \bar r_k(T_k)=0,
\end{array}
\right.
\end{equation}
The convenience of this change of variables lies in the fact that the second equation in \eqref{eq:y,r} can be solved 
autonomously, and then $\bar r_k(t)$ can be made explicit as a function of $\o_k(t)$. We have
\begin{equation}\label{r}
\bar r_k(t)=\eps \int_{T_k}^t e^{-\int_s^t (\o_k(s')+\g)ds'}dw_s
\end{equation}
Through \eqref{r} an explicit formula can be found also for $\bar y_k(t)$ as a function of $\o_k(t)$,
\begin{equation}\label{y}
\bar y_k(t)=\int_{T_k}^t \bar r_k(s) e^{\int_s^t \o_k(s')\, ds'}\;
ds= \eps \int_{T_k}^t \( \int_s^t e^{-\int_s^{s'}(2\o_k(s'')+\ga)\,
ds''}\; ds' \)\; dw_s.
\end{equation}

Finally we define the errors
\begin{equation*}
  \mathcal Y_k(t):=y_k(t)-\bar y_k(t) \quad\quad\quad\quad \text{and} \quad\quad\quad\quad
  \mathcal R_k(t):=r_k(t)-\bar r_k(t)
\end{equation*}

then
\begin{equation}\label{Y}
\mathcal Y_k(t)= \int_{T_k}^t e^{\int_s^t\o_k(s'')\, ds''}\( \int_s^t
e^{-\int_s^{s'}(2\o_k(s'')+\ga)\, ds''}\; ds' \) \varphi(X_k(s),y_k(s))\;
ds
\end{equation}
and
\begin{equation}\label{R}
\mathcal R_k(t)=\int_{T_k}^t e^{-\int_s^t(\o_k(s')+\g)\, ds'}
\varphi(X_k(s),y_k(s))\; ds.
\end{equation}

\subsection*{Remarks}

As long as $|y_k(t)|$ is small enough, thus far from the criticalities, 
\begin{equation*}
r_k(t)=p(t)-\wp_k(X_k(t))-\frac d {dx}\,\wp_k(X_k(t))y_k(t)
= p(t)-\wp_k(x(t))+ \mathcal O(y^2_k(t))
\end{equation*}
and, since we expect $\wp_k(x)$ to be close enough to the
$k$-th heteroclinic orbit $\wp^*_k(x)$, we have
\begin{equation}\label{r:sim}
r_k(t)\simeq p(t)-\wp^*_k(x(t))
\end{equation}

\vskip.3cm

The choice of the variables $(y_k(t),r_k(t))$, thus, turns out to be particularly advantageous. 
 Indeed, with a good choice of the parameter $\eta_\eps$, $r_k(t)$ is not far from $z_k(t)$ just before the criticality
  and from $v_k(t)$ just after it, i.e.
\begin{equation*}
r_k(S_k)\simeq z_k(S_k) \quad \:\text{and}\:\quad
r_{k+1}(T_{k+1})\simeq v_k(T_{k+1})
\end{equation*}
as it is clear from \eqref{z:sim}, \eqref{v:sim} and
\eqref{r:sim}. Under this change of variables, the dynamics far from 
 criticalities becomes ``almost
unidimensional''. Getting in the $k$-th criticality we just need, as
input, the distribution of $z_k(S_k)$ that is provided, unless
small errors, by $r(S_k)$.  Departing from the $k$-th criticality  we get as
output the distribution of $v_k(T_{k+1})|\{z_k(T_{k+1})>0\}$ that
provides the approximated   value of $r_{k+1}(T_{k+1})$. Away from the
criticalities, the fundamental variable is thus $r_k(t)$ and we
can neglect to carefully analyse the behavior of $y_k(t)$. Since the
linearization of $r_k(t)$ has a quite simple form as a function of
$\o_k(t)=\wp_k(X_k(t))$ (see \eqref{r}) everything can be computed
with a good accuracy.

\vskip.5cm

We introduce some of the parameters involved.
Assuming that the ``$k-1$-th criticality has been crossed'', we
will prove in Section \ref{section:main}  that for any $k$, $\bar z_k(S_k)|S_k,T_k$ is a Gaussian r.v. of
standard deviation $\Theta(\sigma_\eps)$ and expected value that is a  $\mathcal O(\tilde \sigma_\eps)$, with
\begin{equation}\label{sigma}
\sigma_\eps:= \eps \; \eta_\eps^{-\frac 1 {1+\theta}}=
\eps^{1-\frac \nu{\theta +1}} \quad  \quad\quad\quad \text{and} \quad\quad\quad\quad
\tilde \sigma_\eps:=\bar \sigma_\eps \,\eta_\eps^{\frac \ga \b
\sqrt{\ga^2 +4 \b}}= \sigma_\eps \; \eps^\theta.
\end{equation}

On the other hand, we will see  that $\bar v_k(T_{k+1})|S_k$
 in both cases (the $k$-th criticality has
been/not been  crossed)   is, for any $k$, a  Gaussian 
r.v. of  standard deviation $\Theta(\eps)$ and expected value
that is a $\mathcal O(\bar \sigma_\eps)$ with
\begin{equation}\label{bsigma}
\bar \sigma_\eps:=\sigma_\eps^{1+\theta}\; \eta_\eps^{-\theta}=
\sigma_\eps \; \eps^{\(1-\frac{2+\theta}{1+\theta}\, \nu\)\theta}
\end{equation}

It is easy to check that, under the condition \eqref{eq:nu}   the following asympthotic relations hold that will be fundamental in our proof
\begin{equation}\label{cond1}
 \sigma_\eps=o(1), \quad \quad 
\bar \sigma_\eps=o(\sigma_\eps)\quad \text{and} \quad
\tilde \sigma_\eps=o(\bar \sigma_\eps) \quad \quad \text{as} \quad \quad \eps \to 0
\end{equation}
and
\begin{equation}\label{cond2}
\eps= o(\eta_\eps^2),\quad \quad
\sigma_\eps=o(\eta_\eps)\quad \text{and}
\quad \eta_\eps^2=o(\tilde \sigma_\eps) \quad \quad \text{as} \quad \quad \eps \to 0
\end{equation}

\vskip.3cm
 The stable region linearization
\eqref{eq:x,p'} is helpful as long as the dispersion around the
heteroclinic solution is smaller than the distance $2k\pi-x(t)$
itself, then we impose $|\bar z_k(S_k)|<2k\pi-x(S_k)=\Theta(\eta_\eps)$,
this provides the condition $\sigma_\eps =o(\eta_\eps)$ and then the upper bound in \eqref{eq:nu}.
\\
On the other hand, the errors due to the linearization in the
critical interval \eqref{eq:x,p'} are of order
$(x(t)-2k\pi)^2= \Theta(\eta_\eps^2)$. We want such an error to be
small  compared with the minimum distance from the heteroclinic
solution, then we impose $(x(t)-2k\pi)^2<|\bar v_k(t)| \wedge |\bar z_k(t)|$ for $t \in [S_k,T_{k+1}]$.
 This yields $\eta_\eps^2 =o(\tilde \sigma_\eps)$ and thus the lower bound in \eqref{eq:nu}.

\section{Proof of the main result}
\label{section:main}

In this Section we prove Theorem \ref{TEO} assuming the estimates
of the errors due to the linearizations obtained in Section
\ref{sect:Err} and the estimates of the variances computed in
Section \ref{section:VAR}.

\vskip1cm

\begin{lemma}\label{lemma:v}

For any fixed $k$, $t \ge 0$,
\begin{equation*}
\bar v_{k}(S_k+t) \:\sim \: \mathbf{Gauss}\: \Big(\mu_v(t),
\sigma_v(t) \Big)
\end{equation*}
with
\begin{equation}\label{sigmav}
\mu_v(t)=\eta_\eps e^{\lam t}, \quad \text{and}\quad
\sigma_v^2(t)= \frac {\eps^2}{2|\lam|}\:\(1-e^{2 \lam t}\)
\end{equation}
$\bar v_k(S_k+t)$, $t\ge 0$ is independent of
$\mathbf{\mathbf{\sigma}}(S_k)$.
\end{lemma}

\vskip.5cm

{\bf Proof.} Recall the definition of $\bar v_k(t)$ in \eqref{v}
then it is clear that $\bar v_k(S_k+t)|S_k$, $t \ge 0$ has a
Gaussian probability law whose average and variance are given by
\begin{equation*}
\E \[\bar v_k(S_k+t)|S_k \]= \eta_\eps e^{\lam t}
\end{equation*}
and
\begin{eqnarray*}
\v[\bar v_k(S_k+t)|S_k]=  \E \[ \:\Big(\bar v_k(S_k+t)-\E
[\bar v_k(S_k+t)|S_k] \Big)^2\: \Big| \: S_k\] \\
 =\eps^2
\E
\[ \: e^{2 \lam(S_k+t)} \(\int_{S_k}^{S_k+t} e^{-\lam s }\;
dw_s \)^2\: \bigg| \: S_k\]  \\
=\eps^2 \E \[\: e^{2 \lam (S_k+t)} \int_{S_k}^{S_k+t} e^{-2 \lam
s} \; ds \:\]= \frac {\eps^2}{2|\lam|}\,(1-e^{2 \lam t}),
\end{eqnarray*}
then follows the result. \qed

\vskip1cm

\begin{lemma}\label{lemma:z''}

We have
\begin{equation}\label{sigmaz}
\bar z_k(S_k+ t)- \bar z_k(S_k) e^{\lap t}\:\sim \:
\mathbf{Gauss}\: \Big(0, \sigma_{z}(t) \Big), \quad\quad
\sigma^2_z(t):=\frac {\eps^2}{2 \lap}\,(e^{2\lap t}-1).
\end{equation}

\end{lemma}
\vskip.5cm

{\bf Proof.} See the proof of Lemma \ref{lemma:v}.\qed

\vskip1cm

Let us define the process
\begin{eqnarray}\label{hatz}
\hat z_k(t):= \hat z_k(S_k) \: e^{\lap (t-S_k)} \;+ \:\eps \:
e^{\lap t}\int_{S_k}^{t} e^{-\lap s}\; dw_s,\nonumber\\
\text{with} \quad \hat z_k(S_k):=P_k(S_k)-\wp_k^*(X_k(S_k))+\bar
r_k(S_k) \label{zS}
\end{eqnarray}
and
\begin{equation}\label{Ek}
\mathcal E_k(t):=\mathcal Z_k(t)+(z_k(S_k)-\hat z_k(S_k))\,
e^{\lap(t-S_k)}
\end{equation}

 then, by  \eqref{zz}, \eqref{V,Z} and \eqref{Ek}
\begin{eqnarray}\label{zk}
\big[z_k(t)\:\big| \:  z_{k-1}(T_k)>0\big]  \sim \hat z_k(t)
+\mathcal E_k(t), \quad \: t \ge S_k.
\end{eqnarray}

\vskip1cm

\begin{lemma}\label{lemma:z}

For $t \ge 0$,
\begin{equation*}
\hat z_{k}(S_k+t)\; |\; \hat z_k(S_k)=z \:\sim \: \mathbf{Gauss}\:
\Big(\mu_{z}(z,t), \sigma_{z}(t) \Big)
\end{equation*}
with
\begin{equation}\label{mu,si}
\mu_{z}(z,t)=z \,e^{\lap t}, \quad \quad  \sigma^2_z(t) \quad
\text{as in} \quad  \eqref{sigmaz}.
\end{equation}

\end{lemma}
\vskip.5cm

{\bf Proof.} See the proof of Lemma \ref{lemma:v}.\qed

\vskip1cm

\begin{lemma}\label{lemma:3}
We have
\begin{equation}\label{A9}
\hat z_k(S_k)\: \big|\: S_k, T_k \: \sim \:
\mathbf{Gauss}\Big(P_k(S_k)-\wp_k^*(X_k(S_k)),
\sigma^2_r(T_k,S_k)\Big)
\end{equation}
with
\begin{equation}\label{sigmar}
\sigma^2_r(T_k,S_k):=\eps^2\int_{T_k}^{S_k}
e^{-2\int_s^t(\o_k(s')+\ga)\,ds'}\; ds,
\end{equation}
$\o_k(t)$ as in \eqref{o}.
\end{lemma}

\vskip.5cm

{\bf Proof.} From the definitions of $\hat z_k(S_k)$ and $\bar
r_k(t)$ in  \eqref{zS} and \eqref{r}, we have
\begin{equation*}
\hat z_k(S_k)= P_k(S_k)-\wp_k^*(X_k(S_k))+\eps \int_{T_k}^{S_k}
e^{-\int_s^{S_k}(\o_k(s')+\ga)\, ds'}\; dw_s
\end{equation*}
thus \eqref{A9} follows. \qed

\vskip1cm

\begin{lemma}\label{lemma:z'}

We have
\begin{equation*}
\hat z_k(S_k+ t)\: \big | \: S_k, T_k \:\sim \: \mathbf{Gauss}\:
\Big(\mu_{z,k}(t), \sigma_{z,k}(t) \Big)
\end{equation*}
with
\begin{eqnarray}\label{sigmazk'}
\mu_{z,k}(t)=e^{\lap t}\,(P_k(S_k)-\wp_k^*(X_k(S_k)))
\end{eqnarray}
and
\begin{eqnarray}\label{sigmazk}
\sigma_{z,k}^2(t)= \sigma^2_r(T_k,S_k)\, e^{2 \lap t}+
\sigma^2_z(t),
\end{eqnarray}
$\sigma^2_z(t)$ as in \eqref{mu,si}.

\end{lemma}

\vskip.5cm

{\bf Proof.} From the definition of $\hat z_k(t)$ and by Lemma
\ref{lemma:3}, we know that $\hat z_k(S_k+ t)$  is the sum of two
processes that, given $S_k,T_k$ have a Gaussian probability law,
thus also $\hat z_k(S_k+ t)\: \big | \: S_k, T_k$ has a Gaussian
law. We have
\begin{equation*}
\E\[\hat z_k(S_k+t)\: \big | \: S_k,T_k\]=\E\[\hat z_k(S_k)\: \big
| \: S_k,T_k\] e^{\lap t}
\end{equation*}
that yields \eqref{sigmazk'}. By the Ito's formula
\begin{equation*}
\hat z_k^2(S_k+t)=\hat z_k^2(S_k)+\int_{S_k}^{S_k+t} (2 \lap \hat
z_k^2(s)+\eps^2)\, ds +2 \eps \int_{S_k}^{S_k + t}\hat z_k(s)
\;dw_s
\end{equation*}
thus the function $f(t):=\E\[\hat z^2_k(S_k +t)\: \big | \:
S_k,T_k\]$ satisfies the equation
\begin{equation*}
\frac d {dt}\, f(t)= 2 \lap f(t) +\eps^2,
\end{equation*}
then
\begin{eqnarray*}
\v\[\hat z_k(S_k +t)\: \big | \: S_k,T_k\]=\E\[\hat z^2_k(S_k
+t)\: \big | \: S_k,T_k\]\\=\E\[\hat z^2_k(S_k)\: \big | \:
S_k,T_k\]e^{2 \lap t} + \frac {\eps^2}{2 \lap}\,(e^{2\lap t}-1)
\end{eqnarray*}
hence \eqref{sigmazk}. \qed

\vskip2cm

For any $\xi>0$ small enough, we define the sets
\begin{equation}\label{Hp2'}
\mathcal K_{k}^{\xi}:=\left\{ (x,p)\;: \:
p-\lap(x-2k\pi)=\eta_\eps, \: |p-\lam(x-2k\pi)|\le \sigma_\eps
\, \eps^{-\xi}\right\}.
\end{equation}
and
\begin{eqnarray}\label{Hp0'}
&& \mathcal H_k^\xi:=\left\{(x,p): \: p= \eta_\eps +\lam(x-2(k-1)\pi), \: |p-\lap(x-2(k-1)\pi)|\le \bar \sigma_\eps \eps^{-\xi}\right\}, \: k \ge 1
\nonumber\\
&&\mathcal H_{0}:=\{(-\pi, \wp_0^*(-\pi))\}
\end{eqnarray}

 \vskip.5cm

In the following propositions we provide some estimates on
expected value and  variance of $\hat z_k(S_k)$.

\vskip1cm
\begin{prop}\label{lemma:4}
There exists $C>0$ such that, for any $\zeta>0$,
$\xi,\eps$ small enough,
\begin{equation}\label{H3}
\mathbf 1_{(x_k,p_k)\in \mathcal H_k^\xi} \;\Pe_{T_k, x_k,p_k}\left\{\sigma_r(T_k,S_k)= \Theta(\sigma_\eps)\right\} \ge
1-e^{-C\eps^{-2\zeta}}
\end{equation}

\end{prop}

 \vskip.5cm

\begin{prop}\label{lemma:err8}
Suppose that $(x_k,p_k) \in \mathcal H_k^\xi$ with $\xi$ small enough, 
\begin{equation}\label{A}
\frac{|P_k(S_k)-\wp^*_k(X_k(S_k))|}{\sigma_r(T_k,S_k)} =\mathcal O(\eps^{\theta})
\end{equation}
Morover there exists $C>0$ such that, for any $\zeta>0$, $\xi,\eps$ small
enough,
\begin{equation}\label{A1}
\mathbf 1_{(x_k,p_k)\in \mathcal
H_k^\xi}\;\Pe_{T_k,x_k,p_k}\left\{ |P_k(S_k)-\wp_k^*(X_k(S_k))|= \mathcal O(\tilde \sigma_\eps)\right\} \ge 1-e^{-C\eps^{-2\zeta}}
\end{equation}
\end{prop}

\vskip.5cm

Propositions \ref{lemma:err8} and \ref{lemma:4} are proved in
Section \ref{sect:Err}.

\vskip1cm

In the following two propositions we give some estimates on the
errors due to the linearizations. In Proposition \ref{err1} we
provide an estimate from below of $T_{k+1}-S_k$.

\vskip.5cm

\begin{prop}\label{prop:err2}

There exists $C>0$ such that, for any 
$\zeta,\eps$ small enough,
\begin{equation}\label{err32'}
\mathbf 1_{(\bar x_k,\bar p_k)\in \mathcal K_{k}^\xi}\, \Pe_{S_k,\bar x_k,\bar p_k}\left\{\sup_{S_k \le t \le T_{k+1}}|\mathcal Z_k(t)|
e^{-\lap(t-S_k)}=\mathcal O(\eta_\eps^2)\right\}\ge 1- e^{-C
\eps^{-2\zeta}}
\end{equation}
and
\begin{equation}\label{err32}
\mathbf 1_{(\bar x_k,\bar p_k)\in \mathcal K_{k}^\xi}\, \Pe_{S_k,\bar x_k,\bar p_k}\left\{\sup_{S_k \le t \le T_{k+1}}|\mathcal V_k(t)|
=\mathcal O(\eta_\eps^2)\right\}\ge 1- e^{-C \eps^{-2\zeta}}
\end{equation}

\end{prop}

\vskip1cm

\begin{prop}\label{prop:err31}
There exists $C>0$ such that, for any $\xi>0$, $\zeta,\eps$ small
enough,
\begin{equation}\label{err31}
\mathbf 1_{(\bar x_k,\bar p_k)\in \mathcal K_k^\xi}\;\Pe_{S_k,\bar x_k,\bar p_k}\left\{
T_{k+1} \ge S_k + \frac 1{\lap}\, \ln\(\,\frac
{\eta_\eps}{\sigma_\eps}\; \eps^{\xi}\)\right\}\ge 1- e^{-C
\eps^{-2\zeta}}
\end{equation}
\end{prop}

\vskip1cm

We will prove Propositions \ref{prop:err}, \ref{prop:err2} and
\ref{prop:err31} in Section \ref{sect:Err}.

\vskip1cm

\begin{prop}\label{prop:err}

There exists $C>0$ such that, for any $\xi>0$, $\zeta,\eps$ small
enough,
\begin{equation}\label{properr-}
\mathbf 1_{(x_k,p_k)\in\mathcal H_k^\xi}
\;\Pe_{T_k,x_k,p_k}\Big\{|z_k(S_k)-\hat z_k(S_k)|=\mathcal O(\eta_\eps^2 \vee \sigma_\eps
\eta_\eps^{-2}\eps^{1-2\zeta})\Big\}\ge 1-e^{-C\eps^{-2\zeta}}
\end{equation}
and
\begin{equation}\label{properr+}
\mathbf 1_{(\bar x_{k-1},\bar p_{k-1})\in\mathcal K^\xi_{k-1}}\;\Pe_{S_{k-1},\bar x_{k-1},\bar p_{k-1}}\left\{ |p(T_{k})-\wp_{k}^*(x(T_{k}))-
\bar v_{k-1}(T_{k})|=\mathcal O(\eta_\eps^{2})\right\}\ge 1- e^{-C \eps^{-2\xi}}
\end{equation}
\end{prop}

\vskip1cm

We will denote by $\Phi(x)$ the function defined by
\begin{equation}\label{Ph}
\Phi(x):=\frac{1}{\sqrt {2 \pi}}\int_{x}^{+\i} e^{-\frac{u^2}2}\,
du
\end{equation}
\vskip.5cm

\begin{lemma}\label{lemma:err9}
There exists $C>0$ such that,  for any 
$\xi,\eps$ small enough, 
\begin{equation}
\mathbf 1_{(x_k,p_k)\in\mathcal H_k^\xi}
\;\Pe_{T_k,x_k,p_k} \left\{|\hat z_k(S_k)|\le  \sigma_\eps \,
\eps^{-\xi}\right\} \ge 1- e^{-C\eps^{-2\xi}}
\end{equation}
\end{lemma}

\vskip.5cm
 {\bf Proof.}
 Let us suppose $(x_k,p_k)\in\mathcal H_k^\xi$ and consider the event
 \begin{equation*}
 \mathbf A_k :=\Big\{ \sigma_r(T_k,S_k)= \Theta(\sigma_\eps) , \;|P_k(S_k)-\wp_k^*(X_k(S_k))|=\mathcal O(\tilde
\sigma_\eps) \Big\}
\end{equation*}
for suitable $c,c',c''>0$, then
\begin{eqnarray}\label{B1}
\Pe_{T_k,x_k,p_k}\left\{|\hat z_k(S_k)|\ge  \sigma_\eps \,
\eps^{-\xi}\right\}=\E_{T_k,x_k,p_k}\[\Pe \left\{|\hat
z_k(S_k)|\ge  \sigma_\eps \, \eps^{-\xi}\: \Big| \: S_k
\right\}\]\\
\le \E_{T_k,x_k,p_k}\[\mathbf 1_{\mathbf A_k}\Pe
\left\{|\hat z_k(S_k)|\ge \sigma_\eps \, \eps^{-\xi}\: \Big| \:
S_k \right\}\]+\Pe_{T_k,x_k,p_k} \left\{\mathbf
A_k^c\right\}\nonumber
\end{eqnarray}
By Propositions \ref{lemma:err8} and \ref{lemma:4},
\begin{equation}\label{B2}
\Pe_{T_k,x_k,p_k} \left\{\mathbf A_k^c\right\}\le e^{-C
\eps^{-2 \xi}}
\end{equation}
for a suitable $C>0$, $\xi$ small enough, and, by Lemma
\ref{lemma:z'}, for any given $T_k$,
\begin{eqnarray*}
\Pe \left\{|\hat z_k(S_k)|\ge \sigma_\eps \, \eps^{-\xi}\: \Big|
\: S_k \right\}= \Phi \(\frac{\sigma_\eps
\eps^{-\xi}-[P_k(S_k)-\wp_k^*(X_k(S_k))]}{\sigma_r(T_k,S_k)}\)\\+\Phi
\(\frac{\sigma_\eps
\eps^{-\xi}-[P_k(S_k)+\wp_k^*(X_k(S_k))]}{\sigma_r(T_k,S_k)}\)
\end{eqnarray*}
where
\begin{eqnarray*}
\mathbf 1_{\mathbf A_k}\; \frac{\sigma_\eps
\eps^{-\xi}\pm[P_k(S_k)-\wp_k^*(X_k(S_k))]}{\sigma_r(T_k,S_k)}\ge
c\, \eps^{-\xi} -c' \, \frac{\tilde \sigma_\eps}{\sigma_\eps}\ge
c''\, \eps^{-\xi}
\end{eqnarray*}
for  suitable $c,c',c''>0$,  since $\tilde
\sigma_\eps/ \sigma_\eps= \eps^\t=o(1)$ as $\eps \to 0$. We have, thus,
\begin{eqnarray}\label{B3}
\mathbf 1_{\mathbf A_k}\,\Pe \left\{|\hat z_k(S_k)|\ge
\sigma_\eps \, \eps^{-\xi}\: \Big| \: S_k\right\}\le
2\Phi\(c'' \eps^{-\xi}\)\le e^{-C \eps^{-2\xi}}
\end{eqnarray}
for some $C>0$, then the result follows from \eqref{B1},
\eqref{B2} and \eqref{B3}. \qed

\vskip1cm

\begin{prop}\label{prop:err10}
There exists $C>0$ such that, for any for any 
$\xi,\eps$ small enough, 
\begin{equation}\label{nH6}
\mathbf 1_{(x_k,p_k)\in\mathcal H_k^{\xi}}
\;\Pe_{T_k,x_k,p_k} \left\{(x(S_k),p(S_k))\in\mathcal K_{k}^{\xi}\right\} \ge
1- e^{-C \eps^{-2\xi}}
\end{equation}

\end{prop}
\vskip.5cm

{\bf Proof.}
 \eqref{nH6} follows
directly from Lemma \ref{lemma:err9} and \eqref{properr-}, since,
by \eqref{cond2}, $\eps=o(\sigma_\eps)$. \qed

\vskip1cm

\begin{prop}\label{prop:err'}
There exists $C>0$ such that, for any $\eps,\xi>0$ small enough,
\begin{equation}\label{A1}
\mathbf 1_{(x_k,p_k)\in\mathcal H_k^{\xi}}
\;\Pe_{T_k,x_k,p_k} \left\{\sup_{S_k \le t \le T_{k+1}}|\mathcal
E_k(t)|\, e^{-\lap(t-S_k)}= \mathcal O(\eta_\eps^2 \vee \sigma_\eps
\eta_\eps^{-2}\eps^{1-2\xi})\right\}\ge 1- e^{-C \eps^{-2\xi}}
\end{equation}

\end{prop}

\vskip.5cm

{\bf Proof.} Let us suppose $(x_k,p_k)\in\mathcal H_k^{\xi}$ and define $\mathbf B_k:=\left\{\sup_{S_k \le t
\le T_{k+1}}|\mathcal Z_k(t)|\, e^{-\lap(t-S_k)}= \mathcal O(\eta_\eps^2)\right\}$, we have
\begin{eqnarray*}
|\Pe_{T_k,x_k,p_k} \left\{\mathbf
B_k^c\right\}-
\Pe\left\{\mathbf B_k^c \; \big| \; (x(S_k),p(S_k))\in\mathcal K_{k}^{\xi}\right\}
|\hskip4.8cm\\\le |\Pe_{T_k,x_k,p_k} \left\{\mathbf B_k^c,
\; (x(S_k),p(S_k))\in\mathcal K_{k}^{\xi}\right\} -\Pe \left\{\mathbf B_k^c\; \big| \; (x(S_k),p(S_k))\in\mathcal K_{k}^{\xi}\right\}| \\+ \Pe_{T_k,x_k,p_k} \left\{(x(S_k),p(S_k))\notin\mathcal K_{k}^{\xi}\right\} 
\\
\le \Pe \left\{\mathbf B_k^c\; \big| \; (x(S_k),p(S_k))\in\mathcal K_{k}^{\xi}\right\} |1-\Pe_{T_k,x_k,p_k}\left\{(x(S_k),p(S_k))\in\mathcal 
K_{k}^{\xi}\right\}| \\+   \Pe_{T_k,x_k,p_k} \left\{(x(S_k),p(S_k))\notin\mathcal K_{k}^{\xi}\right\}
\\
\le 2  \Pe_{T_k,x_k,p_k} \left\{(x(S_k),p(S_k))\notin\mathcal K_{k}^{\xi}\right\}
\end{eqnarray*}
On the other hand,
\begin{equation}
\Pe\left\{\mathbf B_k^c\; \big| \; (x(S_k),p(S_k))\in\mathcal K_{k}^{\xi}\right\}\le \sup_{(\bar x_k,\bar p_k)\in\mathcal K_k^\xi} \Pe_{S_k,\bar x_k,\bar p_k} \left\{\mathbf B_k^c\right\}
\end{equation}

then, from \eqref{err32'} and \eqref{nH6} it follows that there
exists $C>0$ such that, for $\xi,\eps$ small enough,
\begin{equation}\label{nH11}
\Pe_{T_k,x_k,p_k}\left\{\mathbf B_k^c\right\}\le e^{-C
\eps^{-2\xi}}
\end{equation}
thus, recalling that $\mathcal E_k(t)\, e^{-\lap(t-S_k)}=
\mathcal Z_k(t)\,e^{-\lap(t-S_k)}+(z_k(S_k)-\hat z_k(S_k))$,
\eqref{A1}  follows directly from \eqref{properr-} and
\eqref{nH11}.\qed

\vskip1cm

\begin{prop}\label{prop:P1}

There exists $c>0$ such that, for any $\eps>0$ small enough,
\begin{equation*}
\Big |\mathbf 1_{(x_k,p_k)\in\mathcal H_k^{\xi}}
\;\Pe_{T_k,x_k,p_k} \left\{z_k(T_{k+1})\gtrless
0\right\}-\frac 1 2\Big|\le c \eps^{\theta}.
\end{equation*}

\end{prop}

\vskip.5cm

{\bf Proof.} We assume $(x_k,p_k)\in\mathcal H_k^{\xi}$. We define the event
\begin{equation}\label{Bk}
\mathbf C_k:=\left\{\sup_{S_k \le t \le T_{k+1}}|\mathcal
E_k(t)|\, e^{-\lap(t-S_k)}=\mathcal O(\upsilon_\eps^\xi)\right\}.
\end{equation}
with $\upsilon_\eps^\xi:=\eta_\eps^2\vee \sigma_\eps
\eta_\eps^{-2}\eps^{1-2\xi}$.
 By \eqref{zk} we have
\begin{eqnarray*}
\Pe_{T_k,x_k,p_k} \left\{\mathbf C_k,
\;z_k(T_{k+1})>0\right\}=\Pe_{T_k,x_k,p_k} \left\{\mathbf C_k,
\; \hat z_k(T_{k+1})>-\mathcal E_k(T_{k+1})\right\}
\\\le \Pe_{T_k,x_k,p_k} \left\{\hat z_k(T_{k+1})>- c\, \upsilon_\eps \,
e^{\lap(T_{k+1}-S_k)}\right\}
\end{eqnarray*}
For $\mathbf A_k$ as in the Proof of Lemma
\ref{lemma:err9}, we have
\begin{eqnarray}\label{B6}
\Pe_{T_k,x_k,p_k}  \left\{\hat z_k(T_{k+1})>- c\, \upsilon_\eps\,
e^{\lap(T_{k+1}-S_k)}\right\}\hskip3.5cm\\\le  \E_{T_k,x_k,p_k}
\[\mathbf 1_{\mathbf A_k}\Pe\left\{\hat z_k(T_{k+1})>- c\,
\upsilon_\eps\, e^{\lap(T_{k+1}-S_k)}\: \Big | \: S_k\right\}\]\nonumber\\+ \, \Pe_{T_k,x_k,p_k} \left\{\mathbf A_k^c\right\}.\nonumber
\end{eqnarray}
By Lemma \ref{lemma:z'}, for any given $T_k$ we have
\begin{eqnarray}\label{B7}
\Pe\left\{\hat z_k(T_{k+1})>- c\, \upsilon_\eps\,
e^{\lap(T_{k+1}-S_k)}\: \Big | \: S_k\right\}\hskip1cm\\=\Phi\(\frac{-\mu_{z,k}(T_{k+1}-S_k)-c\,
\upsilon_\eps\,
e^{\lap(T_{k+1}-S_k)}}{\sigma_{z,k}(T_{k+1}-S_k)}\)\nonumber
\end{eqnarray}

 where, from \eqref{sigmazk} and \eqref{sigmazk'},
\begin{eqnarray}\label{B8}
\mathbf 1_{\mathbf A_k}\,\frac{\mu_{z,k}(T_{k+1}-S_k)+
c\,\upsilon_\eps\,
e^{\lap(T_{k+1}-S_k)}}{\sigma_{z,k}(T_{k+1}-S_k)}\\\le
 \mathbf 1_{\mathbf A_k}\,\frac{[P_k(S_k)-\wp_k^*(X_k(S_k))]+
c\,\upsilon_\eps}{\sigma_r(T_k,S_k)}\nonumber\\ \le c'\;  \frac
{\tilde \sigma_\eps +\upsilon_\eps }{\sigma_\eps} \le  c'' \;
\eps^\theta\nonumber
\end{eqnarray}

for suitable $c',c''>0$, since, from \eqref{eq:nu}, $\upsilon_\eps
=o(\tilde \sigma_\eps)$. By \eqref{B6}, \eqref{B7}, \eqref{B8} and
\eqref{B2} it follows that
\begin{eqnarray*}
\Pe_{T_k,x_k,p_k} \big\{\mathbf C_k, \;z_k(T_{k+1})>0\big\}\le
\Phi(-c'' \eps^\t)+ e^{-C \eps^{-2\xi}}\le \frac 1 2 + c''' \,
\eps^\t
\end{eqnarray*}

for suitable $C,c'''>0$, hence, by Proposition \ref{prop:err'},
there exist $c,c',C>0$ such that
\begin{eqnarray*}
\Pe_{T_k,x_k,p_k} \big\{z_k(T_{k+1})>0\big\}\le \Pe_{T_k,x_k,p_k} \big\{\mathbf C_k, \;z_k(T_{k+1})>0\big\}+ \Pe_{T_k,x_k,p_k} \big\{\mathbf
C_k^c\big\}\\\le \frac 1 2 + c \, \(\eps^\t + e^{-C
\eps^{-2\xi}}\)\le \frac 1 2 + c' \, \eps^\t.
\end{eqnarray*}

Analogously,
\begin{eqnarray*}
\Pe_{T_k,x_k,p_k} \big\{\mathbf C_k,
\;z_k(T_{k+1})<0\big\}=\Pe_{T_k,x_k,p_k} \big\{\mathbf C_k, \;
\hat z_k(T_{k+1})<-\mathcal E_k(T_{k+1})\big\}
\\ \le\Pe_{T_k,x_k,p_k} \left\{\hat z_k(T_{k+1})< c\, \upsilon_\eps \,
e^{\lap(T_{k+1}-S_k)}\right\} \\
\le \frac 1 2 + e^{-C \eps^{-2\xi}}
\end{eqnarray*}

for some $C>0$, hence
\begin{eqnarray*}
\Pe_{T_k,x_k,p_k} \big\{z_k(T_{k+1})<0\big\}\le \Pe_{T_k,x_k,p_k} \big\{\mathbf C_k, \;z_k(T_{k+1})<0\big\}+ 
\Pe_{T_k,x_k,p_k} \big\{\mathbf C_k^c\big\}\\\le \frac 1 2 +e^{-C'
\eps^{-2\xi}}
\end{eqnarray*}
for some $C'>0$, this concludes the proof of  the Proposition.
\qed

\vskip1cm

\begin{lemma}\label{lemma:J}
There exists $C>0$ such that, for any fixed $\eps,\xi>0$ small enough,
\begin{equation}\label{J}
\mathbf 1_{(x_k,p_k)\in\mathcal H_k^{\xi}}
\;\Pe_{T_k,x_k,p_k}\left\{ T_{k+1} \le S_k +  \frac 1{\lap}\,
\ln\(\,\frac {\eta_\eps}{\sigma_\eps}\; \eps^{-\xi}\)\right\}\ge
1- C \eps^{\xi}
\end{equation}
\end{lemma}
\vskip.5cm {\bf Proof.} We assume $(x_k,p_k)\in\mathcal H_k^{\xi}$ for some $\xi$ small enough. Let $\mathbf C_k$ be as in \eqref{Bk} and define $\tau:= \ln\(\,\eta_\eps
\sigma_\eps^{-1}\; \eps^{-\xi}\)/\lap$, 
we have
\begin{eqnarray*}
\Pe_{T_k,x_k,p_k}\big\{ T_{k+1} \ge S_k +
\tau\big\}=\Pe_{T_k,x_k,p_k}\left\{ \sup_{0\le t \le
\tau}|z_k(S_k+t)|\le \eta_\eps\right\} \\\le\Pe_{T_k,x_k,p_k}
\left\{ \sup_{0\le t \le \tau}|z_k(S_k+t)|\le \eta_\eps, \;
\mathbf C_k\right\}+ \Pe_{T_k,x_k,p_k}\big\{ \mathbf
C_k^c\big\}
\end{eqnarray*}
By \eqref{A1} we know that there exists $C>0$ such that $\Pe_{T_k,x_k,p_k}\left\{ \mathbf C_k^c\right\}\le e^{-C \eps^{-2\xi}}$ for any
$\eps,\xi$ small enough. From the conditions in \eqref{cond2} we have $(\eta_\eps^2 \vee \sigma_\eps
\eta_\eps^{-2}\eps^{1-2\xi})e^{\lap \tau}=o(\eta_\eps)$, thus, by
\eqref{zk} and the definition of $\mathbf C_k$, 
\begin{eqnarray*}
&&\hskip-1cm \Pe_{T_k,x_k,p_k}\left\{ \sup_{0\le t \le
\tau}|z_k(S_k+t)|\le \eta_\eps, \; \mathbf C_k\right\}\\&&\le
\Pe_{T_k,x_k,p_k}\left\{ \sup_{0\le t \le \tau}|\hat
z_k(S_k+t)|\le 2\eta_\eps \right\}\\&&\hskip1cm=\E_{T_k,x_k,p_k}\[\Pe\left\{\sup_{0\le t \le \tau}|\hat z_k(S_k+t)|\le
2\eta_\eps \;\Big|\;S_k \right\}\]
\end{eqnarray*}
It is easy to show from Lemma \ref{lemma:z}, Proposition \ref{lemma:err8} and Proposition \ref{lemma:4} that for any given $T_k$ there exists
$C>0$ such that, for $\eps$ small enough,
\begin{equation*}
\Pe\left\{\sup_{0\le t \le \tau}|\hat z_k(S_k+t)|\le 2\eta_\eps
\;\Big|\;S_k \right\}\le C \eps^\xi
\end{equation*}

then we get \eqref{J}. \qed

\vskip1cm

\begin{coro}\label{coro:G1}
If $(x_k,p_k)\in\mathcal H_k^{\xi}$ for some $\xi>0$ small enough then, for any fixed $\eps>0$ small enough, $k\ge 0$, $T_{k+1}-S_k$ is
$\Pe_{T_k,x_k,p_k}$-a.s. finite.
\end{coro}
\vskip.5cm
 {\bf Proof.} It directly follows from the previous Lemma. \qed

\vskip1cm

\begin{lemma}\label{lemma:A3}
There exists $C>0$ such that, for any $\eps,\xi>0$ small enough,
\begin{equation}\label{A2}
\mathbf 1_{(x_k,p_k)\in\mathcal H_k^{\xi}}
\;\Pe_{T_k,x_k,p_k}\left\{|v_k(T_{k+1})|\ge  \bar \sigma_\eps
\eps^{-\xi}\right\}\le C \eps^\xi
\end{equation}

\end{lemma}

\vskip.5cm

{\bf Proof.}
Assume that $(x_k,p_k)\in\mathcal H_k^{\xi}$, we first prove that there exists $C>0$ such that
\begin{equation}\label{A2'}
\Pe_{T_k,x_k,p_k}\left\{|\bar v_k(T_{k+1})|\ge  \bar \sigma_\eps
\eps^{-\xi}\right\}\le C \eps^\xi
\end{equation}
for any $\eps,\xi$ small enough.

Let us define the event $\mathbf D_k:=\{T_{k+1} \ge S_k
+  \frac 1{\lap}\, \ln\( \eta_\eps \, \eps^{\xi}
\sigma_\eps^{-1}\)\}$, then
\begin{eqnarray*}
\Pe_{T_k,x_k,p_k}\left\{|\bar
v_k(T_{k+1})-\mu_v(T_{k+1}-S_k)|\ge
\eps^{1-\xi}\right\}\hskip4cm\\\le \E_{T_k,x_k,p_k}\[\mathbf
1_{\mathbf D_k}\Pe\left\{|\bar
v_k(T_{k+1})-\mu_v(T_{k+1}-S_k)|\ge \eps^{1-\xi}\: \Big| \:
S_k \right\}\]\\+\Pe_{T_k,x_k,p_k}\{\mathbf D_k^c\}
\end{eqnarray*}
where, by Lemma \ref{lemma:v}, for any fixed $T_{k+1}$,
\begin{eqnarray*}
\Pe\left\{|\bar v_k(T_{k+1})-\mu_v(T_{k+1}-S_k)|\ge \eps^{1-\xi}\:
\Big| \: S_k \right\}\\=2\Phi\(\frac{
\eps^{1-\xi}}{\sigma_v(T_{k+1}-S_k)}\)
\end{eqnarray*}
By \eqref{sigmav} we have
\begin{eqnarray*}
\mathbf 1_{\mathbf D_k}\Phi\(\frac{
\eps^{1-\xi}}{\sigma_v(T_{k+1}-S_k)}\)\le \Phi\(c \eps^{-\xi}\)\le
c' \eps^{-\xi} e^{-c''\eps^{-2\xi}}
\end{eqnarray*}
for some $c,c',c''>0$, moreover, by Lemma \eqref{lemma:J} there exists $C$ such that
\begin{equation}\label{D1}
\Pe_{T_k,x_k,p_k}\{\mathbf D_k^c\}\le C \eps^\xi
\end{equation}
hence there exists $C'>0$ such that
\begin{equation}\label{A2'}
\Pe_{T_k,\bar x,p_k}\Big\{|\bar v_k(T_{k+1})-\mu_v(T_{k+1}-S_k)|\ge
\eps^{1-\xi}\Big\}\le C' \eps^\xi
\end{equation}
 By \eqref{sigmav},
$\mu_v(T_{k+1}-S_k)=\eta_\eps\, e^{\lam(T_{k+1}-S_k)}$, thus,
from \eqref{D1} and the definition of $\bar \sigma_\eps$
\begin{equation}\label{A8}
\Pe_{T_k,\bar x,p_k}\Big\{ \mu_v(T_{k+1}-S_k)\ge \bar \sigma_\eps\,
\eps^{-\xi}\Big\}\le C \eps^\xi
\end{equation}
then \eqref{A2} descends from \eqref{A2'} and
\eqref{A8} since $\eps= o(\bar \sigma_\eps)$. 
We recall now that $v_k(t)= \bar v_k(t)+\mathcal V_k(t)$, then \eqref{A2'} follows 
from \eqref{A2} and \eqref{err32} since $\eta_\eps^2=o(\bar \sigma_\eps)$. 
\qed

\vskip1.5cm

We decompose the set $\mathcal H_k^\xi$ (see the definition in \eqref{Hp0'}) in the two subests: 
\begin{eqnarray}\label{Hp0''}
&& \mathcal H_k^\xi=  \mathcal L_k^\xi \cap \mathcal M_k  \quad \quad \text{for} \quad \: k \ge 1
\end{eqnarray}
with 
\begin{equation}
\mathcal L_k^\xi:=\left\{(x,p): \:|p-\lap(x-2(k-1)\pi)|\le \bar \sigma_\eps \eps^{-\xi}\right\}
\end{equation}
and
\begin{equation}\label{Mk}
\mathcal M_k:=\left\{(x,p): \: p= \eta_\eps +\lam(x-2(k-1)\pi)\right\} 
\end{equation}

and recall that  $\mathcal H_{0}=\{(-\pi, \wp_0^*(-\pi))\}$. 

\vskip1cm

\begin{lemma}\label{lemma:AA1}
Suppose that $(x,p)\in \mathcal H_{k-1}$, then there exist $c,C>0$ such that
\begin{equation}\label{A10}
\Pe_{T_{k-1},x,p}\left\{(x(T_k),p(T_k)) \notin \mathcal L_k^\xi \right\}\le C \eps^\xi
\end{equation}
and
\begin{equation}\label{A11}
\bigg|\Pe_{T_{k-1},x,p}\left\{(x(T_k),p(T_k)) \in \mathcal H_k^\xi\right\}-\frac 1 2 \bigg|\le c \eps^\theta
\end{equation}
for any $\eps$ small enough.
\end{lemma}

\vskip.5cm
{\bf Proof.}
\eqref{A10} follows from \eqref{A2} since $(x(T_k),p(T_k)) \notin \mathcal L_k^\xi$ if and only if $|v_{k-1}(T_{k})|\ge  \bar \sigma_\eps
\eps^{-\xi}$.
On the other hand, from Propositon \ref{prop:P1} it follows that
\begin{equation}\label{A12}
\bigg|\Pe_{T_{k-1},x,p}\left\{(x(T_k),p(T_k)) \in  \mathcal M_k\right\}-\frac 1 2 \bigg|\le c \eps^\theta
\end{equation}
since $(x(T_k),p(T_k)) \in  \mathcal M_k$ if and only if $z_{k-1}(T_k)>0$. We have
\begin{eqnarray*}
&&\Pe_{T_{k-1},x,p}\left\{(x(T_k),p(T_k)) \in \mathcal H_k^\xi\right\}\\
&&= \Pe_{T_{k-1},x,p}\left\{(x(T_k),p(T_k)) \in \mathcal M_k\right\}- \Pe_{T_{k-1},x,p}\left\{(x(T_k),p(T_k)) \in \mathcal M_k \cup (\mathcal L_k^\xi)^c\right\}
\end{eqnarray*}
then, from \eqref{A10}
\begin{eqnarray}\label{A13}
&&\big|\Pe_{T_{k-1},x,p}\left\{(x(T_k),p(T_k)) \in \mathcal H_k^\xi\right\}-\Pe_{T_{k-1},x,p}\left\{(x(T_k),p(T_k)) \in \mathcal M_k\right\}\big|\nonumber\\
&&\le \Pe_{T_{k-1},x,p}\left\{(x(T_k),p(T_k)) \notin  \mathcal L_k^\xi\right\}\le C \eps^\xi
\end{eqnarray}
then \eqref{A11} follows from \eqref{A12} and \eqref{A13} since $\xi$ is arbitrary.
\qed

\vskip1cm

\begin{lemma}\label{lemma:AA2}
Let $(x,p)\in \mathcal H_0$, then there exist $c,C>0$ such that
\begin{equation}\label{A17}
\(\frac 1 2 - c \eps^\theta\)^k -C \eps^\xi\le  \Pe_{T_0,x,p}\left\{(x(T_k),p(T_k)) \in \mathcal H_k^\xi \right\}\le \(\frac 1 2 + c \eps^\theta\)^k +C \eps^\xi
\end{equation}
for any $\eps$ small enough.
\end{lemma}
\vskip.5cm
{\bf Proof.}
We prove the upper bound.
For $(x,p)\in \mathcal H_{k-2}^\xi$ we have
\begin{eqnarray}
&&\Pe_{T_{k-2},x,p}\left\{(x(T_k),p(T_k)) \in \mathcal H_k^\xi \right\}\nonumber \\&&= \E_{T_{k-2},x,p}\[\mathbf 1_{(x,p)\in \mathcal M_{k-1}}\,\Pe_{T_{k-1},x,p}\left\{(x(T_k),p(T_k)) \in \mathcal H_{k}^\xi\right\}\]\nonumber\\
&&\le \E_{T_{k-2},x,p}\[\mathbf 1_{(x,p)\in \mathcal H_{k-1}^\xi}\,\Pe_{T_{k-1},x,p}\left\{(x(T_k),p(T_k)) \in \mathcal H_k^\xi\right\}\]+ \Pe_{T_{k-2},x,p}\left\{(x(T_{k-1}),p(T_{k-1}))\notin \mathcal L_{k-1}\right\} \nonumber\\
&&\hskip5cm \le \(\frac 1 2 +c \eps^{\theta}\) \Pe_{T_{k-2},x,p}\left\{(x(T_{k-1}),p(T_{k-1})) \in \mathcal H_{k-1}^\xi\right\}+C\eps^\xi \nonumber \\
&& \hskip10cm\le \(\frac 1 2 +c \eps^{\theta}\)^2 + C \eps^\xi
\end{eqnarray}
where the two last inequalities follow from \eqref{A10} and \eqref{A11}. By repeating $k$ times this argument we find that, if $(x,p)\in \mathcal H_0$,
\begin{eqnarray}
\Pe_{T_{0},x,p}\left\{(x(T_k),p(T_k)) \in \mathcal H_k^\xi \right\}\le \(\frac 1 2 +c \eps^{\theta}\)^k+ C \eps^\xi \sum_{i=0}^{k-2}\(\frac 1 2 +c \eps^{\theta}\)^i \le \(\frac 1 2 +c \eps^{\theta}\)^k + C' \eps^\xi
\end{eqnarray}
The lower bound follows from the same argument.
\qed

\vskip1cm

\begin{lemma}\label{lemma:AA3}
Let $(x,p)\in \mathcal H_0$, then there exists $C>0$ such that
\begin{equation}\label{A16}
\Pe_{T_0,x,p}\left\{(x(T_k),p(T_k)) \notin \mathcal L_k^\xi \right\}\le C \eps^\xi
\end{equation}
for any $\eps$ small enough.
\end{lemma}

\vskip.5cm
{\bf Proof.}
For $(x,p)\in \mathcal H_{k-2}^\xi$ we have
\begin{eqnarray}
&&\Pe_{T_{k-2},x,p}\left\{(x(T_k),p(T_k)) \notin \mathcal L_k^\xi \right\}=\\
&& = \E_{T_{k-2},x,p}\[\mathbf 1_{(x,p)\in \mathcal M_{k-1}}\,\Pe_{T_{k-1},x,p}\left\{(x(T_k),p(T_k)) \notin \mathcal L_{k}^\xi\right\}\]\nonumber\\
&&\le \E_{T_{k-2},x,p}\[\mathbf 1_{(x,p)\in \mathcal H_{k-1}^\xi}\,\Pe_{T_{k-1},x,p}\left\{(x(T_k),p(T_k)) \notin \mathcal L_k\right\}\]+ \Pe_{T_{k-2},x,p}\left\{(x(T_{k-1}),p(T_{k-1}))\notin \mathcal L_{k-1}\right\} \nonumber\\
&&\hskip5cm \le C \eps^\xi \, \Pe_{T_{k-2},x,p}\left\{(x(T_{k-1}),p(T_{k-1})) \in \mathcal H_{k-1}^\xi \right\}+C\eps^\xi \nonumber\\
&&\hskip9cm \le C\eps^{\xi}\,\[\(\frac 1 2 + c \eps^\theta\)+ 1\] \nonumber
\end{eqnarray}
the last two inequalities descending from \eqref{A10} and \eqref{A11}. By repeating $k$ times this argument we find that, for $(x,p)\in \mathcal H_0$,
\begin{eqnarray}
\Pe_{T_{0},x,p}\left\{(x(T_k),p(T_k)) \notin \mathcal L_k^\xi \right\} \le  C \eps^\xi \; \sum_{i=0}^{k-1}\(\frac 1 2+c \eps^{\theta}\)^i \le C'\eps^\xi
\end{eqnarray}
this concludes the proof of the Lemma. \qed

\vskip1.5cm

{\bf Conclusion of the Proof of Theorem \ref{TEO}.} 
The processes $z_k(t)$ are well defined only for $k \le \mathcal
N$, then we set
\begin{equation*}
z_k(t)\equiv 0  \quad \quad  \text{for} \quad \quad k \ge \mathcal
N +1
\end{equation*}

Let $(x,p)\in \mathcal H_0$, then we prove that there exists $C>0$ such that
\begin{equation}\label{A15}
\bigg|\Pe_{T_0,x,p}\left\{\mathcal N=k\right\}-\(\frac 1 2+c \eps^\theta\)^{k+1}\bigg|\le C \eps^\xi
\end{equation}
for any $\eps$ small enough.
We have
\begin{eqnarray}\label{E0}
 &&\Pe_{T_0,x,p}\left\{\mathcal N=k\right\}=\Pe_{T_0,x,p}\left\{z_k(T_{k+1})<0, \;
z_{k-1}(T_{k})>0, \; ..., \; z_0(T_1)>0\right\}\nonumber \\
&&=\Pe_{T_0,x,p}\left\{z_k(T_{k+1})<0, \;
z_{k-1}(T_{k})>0\right\}\nonumber\\
&& = \Pe_{T_0,x,p}\left\{(x(T_{k+1}),p(T_{k+1}))\notin \mathcal M_{k+1}, (x(T_k),p(T_k))\in \mathcal M_k\right\}\nonumber\\
&& = \E_{T_0,x,p}\[\mathbf 1_{(x,p)\in\mathcal M_k} \, \Pe_{T_k,x,p} \left\{(x(T_{k+1}),p(T_{k+1}))\notin \mathcal M_{k+1}\right\}\]\nonumber \\
&& \le  \E_{T_0,x,p}\[\mathbf 1_{(x,p)\in\mathcal H_k^\xi} \, \Pe_{T_k,x,p}\left\{(x(T_{k+1}),p(T_{k+1}))\notin \mathcal M_{k+1}\right\}\]+
\Pe_{T_0,x,p}\left\{(x(T_k),p(T_k))\notin \mathcal L_k^\xi \right\} \nonumber\\
&& \hskip6cm\le \(\frac 1 2+c \eps^\theta\) \,\Pe_{T_0,x,p}\left\{(x(T_k),p(T_k))\in\mathcal H_k^\xi\right\}+ C \eps^\xi \nonumber\\
&& \hskip8cm\le \(\frac 1 2+c \eps^\theta\) \[\(\frac 1 2+c \eps^\theta\)^k + C \eps^\xi\] +C \eps^\xi \nonumber \\
&& \hskip10cm \le \(\frac 1 2+c \eps^\theta\)^{k+1}+C' \eps^\xi \nonumber
\end{eqnarray}
the last three inequalities descenging from \eqref{A12}, \eqref{A16} and \eqref{A17}.
This yelds the upper bound in \eqref{A15}, the lower bound can be obtained by an analogous argument.
Thus Theorem \ref{TEO} follows from \eqref{A15}. \qed

\vskip1.5cm

{\bf Proof of Proposition \ref{TEO0}.}
For $\xi>0$ we define the set
\begin{equation*}
\mathcal Q^\xi_k:=\left\{(x,p): x\le 2 k \pi
-\frac{\eta_\eps}{(\lap-\lam)}\(1-\frac {\bar \sigma_\eps} {\eta_\eps} \,\eps^{-\xi}\), \: p<-\frac{\lap \eta_\eps}{(\lap-\lam)}
\(1-\frac{|\lam|\bar \sigma_\eps \eps^{-\xi}}{\lap \eta_\eps}\) \right\}
\end{equation*}
 We prove that, for $(x,p)\in \mathcal H_0$,
\begin{equation}\label{BB2}
\lim_{\eps \to 0} \Pe_{T_0,x,p}\left\{(x(T_{\mathcal N+1}),p(T_{\mathcal N+1}))\notin\mathcal Q^\xi_{\mathcal N}\right\}=0
\end{equation}
We have
\begin{eqnarray}
&&\Pe_{T_0,x,p}\Big\{(x(T_{\mathcal N+1}),p(T_{\mathcal N+1}))\notin\mathcal Q^\xi_{\mathcal N}\Big\}\\
&&= 
 \Pe_{T_0,x,p}\Big\{(x(T_{\mathcal N+1}),p(T_{\mathcal N+1}))\notin\mathcal Q^\xi_{\mathcal N}, \:  z_{\mathcal N}(T_{\mathcal N+1})<0, \:
z_{\mathcal N-1}(T_{\mathcal N})>0\Big\}
\\
&& = \E_{T_0,x,p}\[ \mathbf 1_{(x,p) \in \mathcal M_{\mathcal N}} \; \Pe_{T_{\mathcal N},x,p}
\Big\{
(x(T_{\mathcal N+1}),p(T_{\mathcal N+1}))\notin\mathcal Q^\xi_{\mathcal N}, \:  z_{\mathcal N}(T_{{\mathcal N}+1})=-\eta_\eps\Big\}\]
\end{eqnarray}
it is easy to check from the definition of $\mathcal Q^\xi_k$ that 
\begin{equation*}
\Big\{
(x(T_{\mathcal N+1}),p(T_{\mathcal N+1}))\notin\mathcal Q^\xi_{\mathcal N}, \:  z_{\mathcal N}(T_{\mathcal N+1})=-\eta_\eps\Big\}= \Big\{ |v_{\mathcal N}(T_{\mathcal N+1})|\ge \bar \sigma_\eps \, \eps^{-\xi}\Big\}
\end{equation*}
then
\begin{eqnarray}
&&\Pe_{T_0,x,p}\Big\{(x(T_{\mathcal N+1}),p(T_{\mathcal N+1}))\notin\mathcal Q^\xi_{\mathcal N}\Big\}=\E_{T_0,x,p} \[ \mathbf 1_{(x,p) \in \mathcal M_{\mathcal N}} \; \Pe_{T_{\mathcal N},x,p}
\Big\{ |v_{\mathcal N}(T_{\mathcal N+1})|\ge \bar \sigma_\eps \, \eps^{-\xi} \Big\}\]\nonumber\\
&&\le \E_{T_0,x,p} \[ \mathbf 1_{(x,p) \in \mathcal H^\xi_{\mathcal N}} \; \Pe_{T_{\mathcal N},x,p}
\Big\{ |v_{\mathcal N}(T_{\mathcal N+1})|\ge \bar \sigma_\eps \, \eps^{-\xi} \Big\}\]+ \Pe_{T_0,x,p}\left\{ (x(T_{\mathcal N}), p(T_{\mathcal N})
))\notin\mathcal L^\xi_{\mathcal N}\right\}\nonumber
\end{eqnarray}
then \eqref{BB2} follows from \eqref{A2}, \eqref{A16} and \eqref{eq:TEO''}.\qed


\section{Deterministic paths}\label{section:DET}


In this Section we study the qualitative behavior of the orbits of the system \eqref{X,P} lying in a
neighborhood of the heteroclinic path. We recall that  $\wp_k^*(x)$ is  the heteroclinic
orbit defined in $(2(k-1)\pi, 2k\pi)$. We have the following result.

\vskip.5cm

\begin{lemma}\label{lemma:nH3}
Let us fix $\d>0$ small enough, then, for
any   $2(k-1)\pi < x\le 2(k-1)\pi+\d $,
\begin{equation}\label{p'*+}
\wp_k^*(x)=\lap (1+\mathcal O(\d)) (x-2(k-1)\pi)
\end{equation}
whereas,  for any $2k\pi -\d \le x< 2k\pi$
\begin{equation}\label{p'*-}
\wp_k^*(x)=\lam (1+\mathcal O(\d))(x-2k\pi)
\end{equation}

\end{lemma}

\vskip.5cm

{\bf Proof.} It follows directly from \eqref{la}. \qed

\vskip1cm

We denote by $\wp_k(x)$ a generic orbit in the phase plane close
enough to $\wp_k^*(x)$ in $(2(k-1)\pi,2k\pi)$ in the following sense. We fix
$\eta>0$ small enough and define $x_k, x_k'\in (2(k-1)\pi,2k\pi)$, $x_k<x_k'$ such
that
\begin{equation}\label{x01}
x_k-2(k-1)\pi= \Theta(\eta) \quad \text{and}
\quad 2k\pi-x_k'=\Theta(\eta) \quad \quad \text{as} \quad \eta \to 0
\end{equation}
 we suppose
\begin{equation}\label{p0}
|\wp_k(x_k)-\wp_k^*(x_k)|= o(\eta) \quad \quad \text{as} \quad
\eta \to 0
\end{equation}
then, as we will see in the following Lemma,  $\wp_k(x)$ is well
defined in $[x_k, x_k']$.
 \vskip1cm

\begin{lemma}\label{lemma:nH1}
For $x_k,x_k'$ as in \eqref{x01} and $\wp_k(x)$ satisfying
\eqref{p0}, we define $\rho_k(x):=\wp_k(x)-\wp_k^*(x)$. Let
$f_k^\d(x):=\d^{-\frac \b {\lap^2}}\,e^{\int_{\d}^x \frac
{V'(u)}{\wp_k^{*2}(u)}\, du}$ then, for any $\d>\eta$ small enough,
\begin{equation}\label{nH3}
\rho_k(x) =\rho_k(x_k)\,\(\frac{x_k}{x}\)^{\frac \b {\lap^2}}\,
(1+\mathcal O(\d))\quad \quad \quad \text{for}\quad \quad
x_k\le x\le 2(k-1)\pi+\d
\end{equation}
\begin{equation}\label{nH3'}
\rho_k(x)= \rho_k(x_k)\, x_k^{\frac \b {\lap^2}} \;
f_k^\d(x)\;(1+\mathcal O(\d))\quad \quad \quad \quad \quad \quad
\quad \text{for}\quad\quad 2(k-1)\pi+ \d\le x\le 2k\pi-\d
\end{equation}
and
\begin{equation}\label{nH3''}
\rho_k(x) =\rho_k(x_k)\,\frac{x_k^{\frac \b
{\lap^2}}}{(2k\pi-x)^{\frac \b {\lam^2}}}\; f_k^\d(2k\pi-\d)\;
\;(1+\mathcal O(\d))\quad \quad \quad \text{for}\quad \quad
2k\pi-\d\le x\le x_k'
\end{equation}
\end{lemma}

\vskip.5cm

{\bf Proof.}  Because of the periodicity of the dynamics, it is sufficient to prove the result for $k=1$. In order to lighten the notation, we omit the index $1$ in $\wp_1(x), \wp_1^*(x)$, etc. 
\\We define
$x_*:=\inf\{x \ge x_1\, : \: |\rho(x)|\ge \eta\}$, thus, by
\eqref{eq:p},
\begin{equation}\label{nH0}
\rho'(x)= \frac {V'(x)}{\wp^*(x)\wp(x)}= \frac
{V'(x)}{\wp^*(x)(\wp^*(x) +\rho(x))}
\end{equation}
then
\begin{equation}\label{eq:q}
\rho(x)=\rho(x_1) \; e^{\int_{x_1}^x \frac
{V'(u)}{\wp^*(u)(\wp^*(u) +\rho(u))}\, du} \quad \quad \quad
\text{for} \quad \quad x \ge x_1
\end{equation}
hence there exists a function $g(\cdot)$ such that
\begin{equation}\label{eq:q'}
\rho(x)=\rho(x_1) \; e^{(1+g(x))\, \int_{x_1}^x \frac
{V'(u)}{\wp^{*2}(u)}\, du} \quad\quad \quad  \text{and}\quad\quad
\sup_{x_1<x \le x_*}|g(x)|\le c\, \eta
\end{equation}
for some $c>0$. We fix $\d>0$ small enough, $\d>\eta$, then, by
\eqref{be}, for any $k\ge 0$,
\begin{equation}\label{nH1}
V'(x)=-\b (x-2k\pi)(1+\mathcal O(\d)) \quad\quad  \text{for} \quad \quad |x-2k\pi|\le \d
\end{equation}

By \eqref{p'*+} and \eqref{nH1}, there exists $\bar g(x)$, with $\sup_{0\le x\le
\d}|\bar g(x)|= \mathcal O(\d)$ as $\d \to 0$, such that
\begin{equation}\label{F5}
\int_{x_1}^{x} \frac{V'(u)}{\wp^{*2}(u)}\,du= -\frac
{\b}{\lap^2}\; \ln \(\frac {x}{x_1}\)+\bar g(x)\quad \quad
\text{ for}\quad  x_1\le x\le \d
\end{equation}

and, by \eqref{p'*-} and \eqref{nH1}  there exists $\tilde g(x)$, with
$\sup_{2\pi-\d\le x\le 2\pi}|\tilde g(x)|= \mathcal O(\d)$ as $\d \to 0$, such that
\begin{equation}\label{F4}
\int_{2 \pi-\d}^{x} \frac{V'(u)}{\wp^{*2}(u)}\,du= -\frac
{\b}{\lam^2}\; \ln \(\frac {2\pi-x}{\d}\)+\tilde g(x) \quad
\quad \text{for} \quad  2 \pi -\d \le x\le  2\pi
\end{equation}

 Since $x_1 =o(\d)$,
from \eqref{eq:q}, \eqref{F4} and \eqref{F5} we gather
\begin{equation}\label{F6}
\rho(x)=\rho(x_1)\, e^{\bar g(x)(1+g(x))}\,\(\frac
{x_1}{x}\)^{\frac {\b}{\lap^2}\,(1+g(x))} \quad \quad
\text{for} \quad x_1\le x \le \d,
\end{equation}
then, in particular, from \eqref{p0},
\begin{equation}\label{F6'}
\sup_{x_1\le x \le \d\wedge x_*}|\rho(x)|\le
|\rho(x_1)|=o(\eta)
\end{equation}
thus $x_*>\d$ then \eqref{nH3} follows from \eqref{F6}. From
\eqref{nH3} we have, thus
\begin{equation}\label{F7}
\rho(x)= \rho(x_1)\, x_1^{\frac \b {\lap^2}} \; f^\d(x)
\;(1+\mathcal O(\d))\quad \quad \text{for} \quad \d \le x \le
(2\pi-\d)\wedge x_*
\end{equation}
Since $f^\d(x)$ does not depend on $\eta$,  we have in particular
that $\sup_{\d\le x \le 2\pi-\d}|\rho(x)|=o(\eta)$, hence
$x_*>2\pi-\d$, thus \eqref{nH3'} follows.
 Finally
\begin{equation}\label{F8'}
\rho(x)=  \rho(2\pi-\d)\, e^{\tilde
g(x)(1+g(x))}\,(2\pi-x)^{-\frac {\b}{\lam^{2}}\,(1+g(x))}\quad
\text{for} \quad 2 \pi -\d \le x < 2 \pi
\end{equation}
thus
\begin{eqnarray}
\sup_{2\pi-\d\le x\le x_1'\wedge x_*}|\rho(x)|=
|\rho(x_1')|=|\rho(2\pi-\d)|\, \mathcal O\(\eta^{-\frac
{\b}{\lam^{2}}}\)\nonumber
\\=|\rho(x_1)|\, \mathcal O\(\eta^{\b\(\frac
{1}{\lap^{2}}-\frac {1}{\lam^{2}}\)}\)=o(\eta) \quad \quad \text{as}
\quad \quad  \eta \to 0\label{F8''}
\end{eqnarray}
since $\frac {\b}{\lap^{2}}>\frac {\b}{\lam^{2}}$. We have, thus,
$x_*>x_1'$, then \eqref{nH3''} follows from \eqref{F7} and
\eqref{F8'}. \qed

\vskip1cm

\begin{remark1}
Notice that
\begin{equation}\label{rem1}
\rho_k(x)= o(\eta) \quad\quad\quad\quad \text{for any} \quad \quad
 x_k\le x\le  x_k'
\end{equation}
and, by \eqref{F8''},
\begin{equation}\label{rem1}
\rho(x_k')= |\rho(x_k)| \; \mathcal
O\(\eta^{\frac{2+\theta}{1+\theta}\,\theta}\)\quad \quad \text{as}
\quad \quad  \eta \to 0
\end{equation}
since $\frac {\b}{\lap^{2}}-\frac {\b}{\lam^{2}}=\frac \ga \b
\sqrt{\ga^2+4\b}= \frac{2+\theta}{1+\theta}\,\theta$.
\end{remark1}

\vskip1cm

\begin{lemma}\label{lemma:derivata}
For $x_k,  x_k'$ as in \eqref{x01} and $\wp_k(x)$ satisfying \eqref{p0},
\begin{equation}\label{derivata}
 \sup_{x_k\le x \le  x_k'}\bigg|\frac{d}{dx}\,\wp_k(x)\bigg|=\mathcal O(1) \quad \quad \text{as} \quad \eta \to 0
\end{equation}
\end{lemma}

{\bf Proof.}
$\wp_k(x)$ verifies the equation
\begin{equation}\label{der1}
\frac{d}{dx}\, \wp_k(x)=- \frac{V'(x)}{\wp_k(x)}-\gamma
\end{equation}
then it is sufficient to prove that $V'(x)/\wp_k(x)$ is uniformly bounded 
in $\eta$ in a neighborhood of $x_k$ and $x_k'$.
Thus the result easily follows by expanding $V'(x)$ and $\wp_k(x)$ in a neighborhood of $x_k$
and $x_k'$  and by using Lemma \ref{lemma:nH3} and Lemma \ref{lemma:nH1}. \qed

\vskip1cm

By similar arguments can be proved the following Lemma.

\begin{lemma}\label{lemma:derivata'}
For $x_k,  x_k'$ as in \eqref{x01} and $\wp_k(x)$ satisfying \eqref{p0} we have
\begin{equation}\label{F10}
\Big|\frac d{dx}\, \wp_k(x)-\frac d {dx}\,\wp_k^*(x)\Big| \le
|\wp_k(x_k)-\wp_k^*(x_k)|\; \eta^{-1} \quad \text{for} \quad
 x_k\le x \le  x_k'
\end{equation}
\end{lemma}

\vskip.5cm

We define, now, the functionals
\begin{equation}
\Sigma_r^{(n)}[\wp, \bar x](x):= \int_{\bar x}^x \frac{du}{\wp(u)}
\: e^{n\int_u^x \frac {V'(u')}{\wp^2(u')}\, du'}
\end{equation}
and
\begin{equation}
\Sigma_y^{(n)}[\wp,\bar x](x):=\wp^n(x)\; \int_{\bar x}^x \;
\frac{du}{\wp(u)}
\[\int_u^x  \frac{du'}{\wp^2(u')}\;
e^{\int_{u}^{u'} \frac {V'(u'')}{\wp^2(u'')}\, du''}\]^n
\end{equation}
$n\in\mathbb{N}$,  $x \ge \bar x$. In the rest of the Section we provide 
some estimates on $\Sigma_r^{(n)}[\wp, \bar x](\cdot)$ and $\Sigma_y^{(n)}[\wp,\bar x](\cdot)$ that 
are fundamental for the study of the variances in Section \ref{section:VAR}.

\vskip.5cm
For $\eta,  x_k,  x_k', \wp_k(x)$ verifying the conditions in \eqref{x01} and \eqref{p0}, we prove the following Lemma.
\vskip.5cm

\begin{lemma}
For any $k\ge 1, n \in \mathbb{N}$ we have
\begin{equation}\label{f*}
\sup_{x_k\le x \le
 x_k'}\frac{\Sigma_r^{(n)}[\wp_k, x_k](x)}{\rho_k(x)^n}\;\rho_k(x_k)^n\, x_k^{n(1+\theta)}=
\mathcal O(1)\quad \quad \text{as} \quad \quad \eta \to 0
\end{equation}
moreover
\begin{equation}\label{f*'}
\Sigma_r^{(n)}[\wp_k, x_k](x_k')\; \frac{\rho_k(x_k)^n\, x_k^{n(1+\theta)}}{\rho_k(x_k')^n} = \Theta\(1\)\quad \quad \text{as} \quad \quad \eta \to 0
\end{equation}
\end{lemma}
\vskip.5cm

{\bf Proof.}  As before, we  prove the statement  for $k=1$ and omit the index $1$ in the notation. \\At first we prove that
\begin{equation}\label{f}
\bigg| \frac{\Sigma_r^{(n)}[\wp, x_1](x)}{\rho(x)^n\,
F_n[\wp, x_1](x)}-1\bigg|= o(\eta) \quad \quad \text{with} \quad
\quad F_n[\wp, x_1](x):=\int_{x_1}^x
\frac{du}{\wp(u)\rho(u)^n}
\end{equation}
From \eqref{nH0} we have
\begin{equation}\label{nH00}
\rho'(x)= \frac {V'(x)}{\wp(x)(\wp(x) -\rho(x))}
\end{equation}
then
\begin{equation}\label{eq:qq}
\rho(x)=\rho(u) \; e^{\int_{u}^x \frac {V'(u')}{\wp(u')(\wp(u')
-\rho(u'))}\, du'} \quad \quad \quad \text{for any} \quad \quad
 x_1 \le u \le x
\end{equation}
thus, from \eqref{rem1} there exists $\hat g(x)$ such that
\begin{equation}\label{eq:qq'}
\rho(x)=\rho(u) \; e^{(1+\hat g(x))\, \int_{u}^x \frac
{V'(u')}{\wp^{2}(u')}\, du'} \quad\quad \quad  \text{and}\quad\quad
\sup_{x_1 \le x \le x_1'}|\hat g(x)|=o(\eta)
\end{equation}
it follows that
\begin{equation}\label{eq:qqq}
\sup_{x_1 \le u \le x \le x_1'}\bigg |  \frac
{\rho(u)^n}{\rho(x)^n} \;e^{n \int_u^x \frac {V'(u')}{\wp(u')^2}\,
du'}-1 \bigg|=o(\eta)
\end{equation}
this yields \eqref{f}.

\vskip.3cm

Let $\d$ be small enough, $x \in [x_1,\d]$, then by
\eqref{p'*+} and \eqref{nH3},
\begin{eqnarray}
F_n[\wp,x_1](x)=(1+\mathcal O(\d))\, \frac
{x_1^{-\frac{n\b}{\lap^2}}} {\lap \rho(x_1)^n }\;
\int_{x_1}^x u^{-1+\frac {n\b}{\lap^2}}\, du \\ = (1+\mathcal
O(\d))\, \frac {\lap} {n \b \rho(x_1)^n}\;\[\(\frac x
{x_1}\)^{\frac {n \b}{\lap^2}}-1\]\quad \quad \text{for} \quad
x_1 \le x \le \d\label{f1}
\end{eqnarray}
Let, now, $x \in [\d, 2\pi-\d]$, then $F_n[\wp, x_1](\d)=F_n[\wp,
x_1](\d)+F_n[\wp, \d](x)$ with, by \eqref{nH3'} and
\eqref{rem1},
\begin{equation}\label{f2}
F_n[\wp,\d](x)= (1+\mathcal O(\d))\,\frac {x_1^{-\frac {n\b}
{\lap^2}}}{\rho(x_1)^n}\;\int_{\d}^{x} \frac {du}{\wp^*(u)\;
f^\d(u)^n} \quad \quad \text{for} \quad \d\le x \le 2\pi-\d
\end{equation}
with $f^\d(x)$ as in Lemma \ref{lemma:nH1}. Finally, for $x \in
[2\pi-\d,x_1']$, $F_n[\wp, x_1](x)=F_n[\wp,
x_1](2\pi-\d)+F_n[\wp, 2\pi-\d](x)$, with,  by \eqref{p'*-} and
\eqref{nH3''},
\begin{eqnarray}
F_n[\wp,2\pi-\d](x)= (1+\mathcal O(\d))\, \,\frac {x_1^{-\frac
{n\b}
{\lap^2}}}{|\lam|\,f^\d(2\pi-\d)^n\,\rho(x_1)^n}\;\int_{2\pi-x}^\d
u^{-1+\frac{n \b}{\lam^2}}\, du\\
=(1+\mathcal O(\d))\, \,\frac {|\lam|\,x_1^{-\frac {n\b}
{\lap^2}}}{n \b \,f^\d(2\pi-\d)^n\,\rho(x_1)^n}\;\[\d^{\frac
{n\b} {\lam^2}}-(2\pi-x)^{\frac {n\b} {\lam^2}}\]\quad \quad
\text{for} \quad 2\pi-\d\le x \le x_1'\label{f3}
\end{eqnarray}
From \eqref{f1}, \eqref{f2} and \eqref{f3} we have that, for any
$x \in [x_1,x_1']$,
\begin{equation}\label{f4}
F_n[\wp,x_1](x)\,\rho(x_1)^n\,x_1^{\frac {n\b}
{\lap^2}}=\mathcal O(1)  \quad \quad \text{as} \quad \quad \eta
\to 0
\end{equation}
and, in particular,
\begin{equation}\label{f5}
\lim_{\eta \to 0}
F_n[\wp,x_1](x_1')\,\rho(x_1)^n\,x_1^{\frac {n\b}
{\lap^2}}=(1+\mathcal O(\d))\,\(\frac {\lap\, \d^{\frac {n
\b}{\lap^2}}} {n \b }+\int_{\d}^{2\pi-\d} \frac {du}{\wp^*(u)\;
f^\d(u)^n}+\frac {|\lam|\;\d^{\frac {n\b} {\lam^2}}}{n \b
\,f^\d(2\pi-\d)^n}\)
\end{equation}

Hence, by \eqref{f4} and \eqref{f}, for any $x \in
[x_1,x_1']$,
\begin{equation}
\frac{\Sigma_r^{(n)}[\wp,x_1](x)}{\rho(x)^n}\;
\,\rho(x_1)^n\,x_1^{\frac {n\b} {\lap^2}}= \mathcal O(1)
\quad \quad \text{as} \quad \quad \eta \to 0
\end{equation}
then \eqref{f*} follows since $\b /\lap^2=1+\theta$. From
\eqref{f5} and the definition of $f^\d(x)$ it is clear that the
limit
\begin{eqnarray}\label{f6}
\lim_{\eta \to 0}
F_n[\wp,x_1](x_1')\,\rho(x_1)^n\,x_1^{\frac {n\b}
{\lap^2}}= \\
\lim_{\d \to 0}\(\d^{\frac {n\b} {\lap^2}}\int_\d^{2\pi-\d}\frac
{du}{\wp^*(u)}\; e^{-n\int_\d^u \frac{V'(u')}{\wp^{*2}(u')}\, du'} +
\frac{|\lam|\d^{n\b (\lap^{-2}+\lam^{-2})}}{n\b}\;
e^{-n\int_\d^{2\pi-\d} \frac{V'(u')}{\wp^{*2}(u')}\, du'}\)
\end{eqnarray}
must be finite and strictly positive, thus \eqref{f*'} follows
from \eqref{f6} and \eqref{f}. \qed

\vskip1cm

\begin{coro}
From \eqref{f*} and \eqref{x01} it follows that
\begin{equation}\label{f**}
\sup_{x_k\le x \le  x_k'}\Sigma_r^{(n)}[\wp_k, x_k](x)=\mathcal O\(\eta^{-\frac{n}{1+\theta}}\)\quad \quad \text{as} \quad \quad \eta \to 0
\end{equation}
and
\begin{equation}\label{g**}
 \Sigma_r^{(n)}[\wp_k, x_k](x_k')= \Theta\(\eta^{-\frac{n}{1+\theta}}\) \quad \quad \text{as} \quad \quad \eta \to 0
\end{equation}
\end{coro}
\vskip.5cm
{\bf Proof.}
The result follows since, from Lemma \ref{lemma:nH1},
\begin{equation}
\sup_{x_k\le x \le  x_k'}\frac{\rho_k(x)}{\rho_k(x_k)}=\frac{\rho_k(x_k')}{\rho_k(x_k)}
\end{equation}
where
\begin{equation}\label{rem1'}
\frac{\rho_k(x_k')}{\rho_k(x_k)}= \Theta \(\eta^{\frac{2+\theta}{1+\theta}\, \theta}\)
\end{equation}
since $\frac{\b}{\lap^2}-\frac{\b}{\lam^2}=\frac{\ga}{\b}\sqrt{\ga^2+4\b}=\frac{2+\theta}{1+\theta}\, \theta$. \qed

\vskip1cm

\begin{lemma}
We have
\begin{equation}\label{q}
\sup_{x_k\le x \le
 x_k'}\big|\Sigma_y^{(n)}[\wp_k, x_k](x)\big|= \mathcal
O(\eta^{-n}) \quad \quad \text{as} \quad \quad \eta \to 0
\end{equation}
\end{lemma}

\vskip.5cm

{\bf Proof.} We prove the result for $k=1$ and omit the index $1$. From \eqref{eq:qqq}, it follows that
\begin{equation}\label{q0}
\sup_{x_1\le x \le
 x_1'}\Bigg|\frac{\Sigma_y^{(n)}[\wp, x_1]}{\wp(x)^n}
\left\{\int_{x_1}^{x} \frac{B(u,x)^n \,
du}{\wp(u)\rho(u)^n}\right\}^{-1}-1 \Bigg|=o(\eta) \quad \quad
\text{as} \quad \quad \eta \to 0
\end{equation}

with
\begin{equation}
B(u,x):=\int_u^x \frac{\rho(u')\, du'}{\wp(u')^2}
\end{equation}

By the same techniques used in the proof of the previous Lemma it
is possible to show that there exists $c>0$ such that
\begin{equation}
B(u,x)\le c \, \rho(x_1)\, \eta^{\frac \b
{\lap^2}}\[u^{-(1+\frac{\b}{\lap^2})}\,\mathbf 1_{x_1\le u
\le \d} + \mathbf 1_{\d\le u \le 2\pi-\d}+
(2\pi-u)^{-(1+\frac{\b}{\lam^2})}\,\mathbf 1_{2\pi-\d\le u \le
 x_1'}\]
\end{equation}
for any $\eta$ small enough, then, by Lemma \ref{lemma:nH1}, there
exists $c>0$ such that
\begin{equation}
\frac{B(u,x)}{\rho(u)}\le c \,
\[u^{-1}\,\mathbf 1_{x_1\le u \le \d} +
\mathbf 1_{\d\le u \le 2\pi-\d}+ (2\pi-u)^{-1}\,\mathbf
1_{2\pi-\d\le u \le x_1'}\]
\end{equation}
hence
\begin{equation}\label{q1}
\sup_{x_1\le x \le x_1'} \int_{x_1}^{x} \frac{B(u,x)^n \,
du}{\wp(u)\rho(u)^n}= \mathcal O(\eta^{-n})  \quad \quad \text{as}
\quad \quad \eta \to 0
\end{equation}
thus \eqref{q} follows from \eqref{q0} and \eqref{q1}. \qed

\section{Estimates of the variances}
\label{section:VAR}

We recall that  $(X_k(t),P_k(t)), \; T_k \le t \le S_k$ is the solution of the problem
\begin{equation}
\left\{
\begin{array}{ll}
\dot {X}_k=P_k & X_k(T_k)= x_k\\
\dot {P}_k=-\gamma P_k-V'(X_k) & P_k(T_k)= p_k
\end{array}
\right.
\end{equation}
$\wp_k(x)$ is the related orbit in $[2(k-1)\pi, 2k \pi]$, i.e. the path such that $\wp_k(X_k(t))=P_k(t)$ for $T_k \le t \le S_k$ and $\o_k(t)=\frac d {dx} \wp_k(X_k(t))$.

\vskip.4cm
In this Section we provide some estimates on the variances of the processes $\bar
y_k(t)$  and $\bar r_k(t)$. The two following  Lemmas follow
directly from their definitions in \eqref{y} and \eqref{r}.
\vskip.4cm

\begin{lemma}\label{lemma:r}
Let
\begin{equation}\label{sigmar'}
\sigma^2_r(T_k,t):=\eps^2\int_{T_k}^{t}
e^{-2\int_s^t(\o_k(s')+\ga)\,ds'} \; ds
\end{equation}
then
\begin{equation*}
 \bar r_k(t)\: | \: T_k \quad \sim \quad
 \mathbf{Gauss}\(0,\sigma_r(T_k,t)\), \quad \quad t \ge T_k
\end{equation*}
\end{lemma}

\vskip.4cm

\begin{lemma}\label{lemma:y}
Let
\begin{eqnarray}\label{sigmay}
\sigma^2_y(t,T_k):=\eps^2 \int_{T_k}^{t} e^{2\int_s^t \o_k(s'')\,
ds''}\[\int_s^t e^{-\int_s^{s'} (2\o_k(s'')+\ga)\,ds''}\; ds'\]^2 \; ds
\end{eqnarray}

then
\begin{equation*}
 \bar y_k(t)\: | \: T_k \quad \sim \quad \mathbf{Gauss} \: \(0,
 \sigma_y(T_k,t)\), \quad \quad t \ge T_k
\end{equation*}
\end{lemma}

\vskip.4cm

Let  $\eta_\eps>0$ be as in Section \ref{sect:Intro}, we recall that
\begin{equation}
\mathcal H_k^\xi=\left\{(x,p): \: p= \eta_\eps +\lam(x-2(k-1)\pi), \: |p-\lap(x-2(k-1)\pi)|\le \bar \sigma_\eps \eps^{-\xi}\right\}, \quad \quad \text{for} \quad k \ge 1
\end{equation}
and $\mathcal H_0=\{(-\pi,\wp_0^*(-\pi))\}$, and define the stopping time
\begin{equation}\label{S}
\bar S_k:=\inf\left\{t \ge T_k: X_k(t)\ge 2k \pi -
\frac{\eta_\eps}{2(\lap-\lam)}\right\}
\end{equation}
We denote by $x_k':=X(\bar S_k)$, then
\begin{equation}\label{xk'}
2k\pi-x_k'=\frac{\eta_\eps}{2(\lap-\lam)}
\end{equation}
and $x_k\le X_k(t)\le x_k'$ for $T_k \le t \le \bar S_k$.
\vskip.5cm
In the following Lemma we prove that, if $(x_k, p_k) \in \mathcal H^\xi_k$, then $x_k$, $x_k'$ and $\wp_k(\cdot)$ satisfy the
conditions \eqref{x01} and \eqref{p0}  in $[2(k-1)\pi, 2k\pi]$ in the following sense.
\vskip.5cm

\begin{lemma}\label{lemma:cond}
Let $(x_k,p_k) \in \mathcal H^\xi_k$,  with $\xi$ small enough, then 
\begin{equation}\label{x01'}
x_k-2(k-1)\pi= \Theta\(\eta_\eps\)  \quad 
\quad \text{and}\quad \quad 2 k\pi-x_k'= \Theta\(\eta_\eps\)
 \quad \quad \text{as} \quad \quad \eps \to 0
\end{equation}
moreover 
\begin{equation}\label{p0'}
|\wp_k(x_k)-\wp_k^*(x_k)|=
o(\eta_\eps) \quad \quad \text{as} \quad \quad \eps \to 0
\end{equation}

\end{lemma}
\vskip.5cm

{\bf Proof.}
\eqref{x01'} follows directly from the definition of
$\mathcal H^\xi_k$ and from \eqref{xk'},
whereas \eqref{p0'} is verified since
\begin{eqnarray}
&&|\wp_k(x_k)-\wp_k^*(x_k)|=|p_k-\wp_k^*(x_k)|\nonumber \\
&&\le |p_k-\lap(x_k-2(k-1)\pi)|+|\lap(x_k-2(k-1)\pi)-\wp_k^*(x_k)| \nonumber \\
&&\le \bar \sigma_\eps \eps^{-\xi}+ c \eta_\eps^2\le 2 \bar \sigma_\eps \eps^{-\xi}\label{p0''}
\end{eqnarray}
where the last inequality follows from the definition of $\mathcal{H}_k^\xi$, \eqref{p'*+} and the left hand side of \eqref{x01'}. Hence \eqref{p0'} holds if $\xi$ is small enough, since, from \eqref{cond2}, $\bar{\sigma}_\eps =o(\eta_\eps)$ as $\eps \to 0$.
\qed
 \vskip1cm

\begin{lemma}\label{lemma:cond'}
Let $(x_k,p_k) \in \mathcal H^\xi_k$,  with $\xi$ small enough, then 
\begin{equation}\label{x02}
\wp_k(x_k')=\lam(x_k'-2k\pi)+\mathcal O\(\tilde \sigma_\eps \, \eps^{-\xi}\) \quad \quad \text{as} \quad \quad \eps \to 0
\end{equation}

\end{lemma}
\vskip.5cm
{\bf Proof.} From  \eqref{rem1} and \eqref{p0''}  we have
\begin{equation}
|\wp_k(x_k')-\wp_k^*(x_k')|\le c \,\eta_\eps^{\frac{2+\theta}{1+\theta}\, \theta} \bar \sigma_\eps \eps^{-\xi}= \mathcal O\(\tilde \sigma_\eps \eps^{-\xi}\)\quad \quad \text{as} \quad \quad \eps \to 0
\end{equation}
then the result follows from \eqref{p'*-} since, from \eqref{cond2},  $\eta_\eps^2=o(\tilde \sigma_\eps)$ as $\eps \to 0$. \qed
\vskip1cm

\begin{remark1}
As a consequence of  Lemma \ref{lemma:cond} and the considerations done in Section 4.1,   $\wp_k(\cdot)$ is well defined in $[x_k,x_k']$.
\end{remark1}

\vskip1cm

\begin{prop}\label{prop:sigmar}
Let $(x_k,p_k)\in \mathcal H^\xi_k$, $\xi$ small enough, then
\begin{equation}\label{err9} \sup_{T_k \le t \le
\bar S_k}\sigma^2_r(T_k,t)= \mathcal O(\sigma_\eps^2) \quad \quad
\text{as} \quad \eps \to 0 \quad \quad \Pe_{T_k, x_k,p_k} \: a.s.
\end{equation}
\end{prop}

 \vskip.5cm

{\bf Proof.} 
 We have
\begin{eqnarray}\label{SSS}
\sigma_r^2(T_k,t) =\eps^2 \int_{x_k}^{X_k(t)}e^{-2
\int_u^{X_k(t)} \frac{\wp_k'(u')+\ga}{\wp_k(u')}\, du'}
\;\frac{du}{\wp_k(u)}= \eps^2 \Sigma_r^{(2)}[\wp_k,
x_k](X_k(t))
\end{eqnarray}
where the second identity follows from \eqref{eq:p}, then
\begin{equation}
\sup_{T_k \le t \le \bar S_k}\sigma_r^2(T_k,t)=\eps^2 \sup_{x_k\le x \le x_k'} \Sigma_r^{(2)}[\wp_k,
x_k](x)= \eps^2 \mathcal O\big(\eta_\eps^{-\frac 2 {1+\theta}}\big)
\end{equation}
where the last equivalence follows from \eqref{f**}.
Then \eqref{err9} follows from the definition of $\sigma_\eps$ in
\eqref{sigma}. \qed

\vskip1cm

\begin{prop}\label{prop:sigmay}
Let $(x_k,p_k)\in \mathcal H^\xi_k$, $\xi$ small enough, then
\begin{equation}\label{err1}
\sup_{T_k\le t \le \bar S_k} \sigma_y^2(T_k,t) = \mathcal O\(\eps^2\,
\eta_\eps^{-2}\) \quad \quad \Pe_{T_k,x_k,p_k} \:  a.s.
\end{equation}
\end{prop}

\vskip.5cm
{\bf Proof.}
We have
\begin{eqnarray}
&&\sigma_y^2(T_k,t) = \eps^2 \int_{T_k}^t e^{2\int_{X_k(s)}^{X_k(t)} \frac{\wp_k'(u')}{\wp_k(u')}du'}\[ \int_s^t e^{-\int_{X_k(s)}^{X_k(s')} \frac{2\wp_k'(u')+\gamma}{\wp_k(u')} du'} \, ds'\]^2\; ds \\
&&= \wp_k^2(X_k(t)) \int_{x_k}^{X_k(t)} \wp_k(u)\[\int_u^{X_k(t)} \frac {du'}{\wp_k^3(u')} e^{-\int_u^{u'} \frac \gamma {\wp_k(u'')} \, du''}\]^2\; du \\
&&= \eps^2 \Sigma_y^{(2)}[\wp_k, x_k](X_k(t))
\end{eqnarray}
thus
\begin{equation}
\sup_{T_k\le t \le \bar S_k} \sigma_y^2(T_k,t) = \eps^2 \sup_{x_k\le x \le x_k'} \Sigma_y^{(2)}[\wp_k, x_k](x)= \eps^2 \, \mathcal O\(\eta_\eps^{-2}\)
\end{equation}
where the last equivalence follows from \eqref{q}. \qed

\vskip1cm

\begin{lemma}\label{omega}
Let $(x_k,p_k)\in \mathcal H^\xi_k$, $\xi$ small enough, then there exists $C>0$ such that
\begin{equation}
\sup_{\eps<1}\sup_{T_k \le t \le \bar S_k}|\omega_k(t)|\le C
\end{equation}

\end{lemma}
\vskip.5cm
{\bf Proof.} It is a direct consequence of Lemma \ref{lemma:derivata}, since
\begin{equation}
\sup_{T_k \le t \le \bar S_k}|\omega_k(t)|= \sup_{x_k\le x \le x_k'}\bigg|\frac{d}{dx}\,\wp_k(x)\bigg|
\end{equation}
\qed

\vskip1cm

\begin{lemma}\label{H}
Let 
\begin{equation}
H_r(T_k,t):=\int_{T_k}^{t}
e^{-\int_s^t(\o_k(s')+\ga)\,ds'} \; ds
\end{equation}
and  
\begin{equation}
H_y(t,T_k):=\int_{T_k}^{t} e^{\int_s^t \o_k(s'')\,
ds''}\[\int_s^t e^{-\int_s^{s'} (2\o_k(s'')+\ga)\,ds''}\; ds'\] \; ds
\end{equation}
then, for $(x_k,p_k)\in \mathcal H^\xi_k$, $\xi$ small enough, we have
\begin{equation}\label{Hr}
\sup_{T_k \le t \le \bar S_k} H_r(T_k,t)= \mathcal O\(\eta_\eps^{-\frac{1}{1+\theta}}\)
\end{equation}
and
\begin{equation}\label{Hy}
\sup_{T_k \le t \le \bar S_k} H_y(T_k,t)= \mathcal O\(\eta_\eps^{-1}\)
\end{equation}
\end{lemma}
\vskip.5cm
{\bf Proof.}
It follows directly from \eqref{f**} and \eqref{q} since
\begin{equation}
H_r(T_k,t)=\Sigma_r^{(1)}[\wp_k, x_k](X_k(t)) \quad \quad \text{and} \quad \quad H_y(T_k,t)=\Sigma_y^{(1)}[\wp_k, x_k](X_k(t))
\end{equation}
\qed

\section{Estimate of Errors}
\label{sect:Err}

In this Section we prove Propositions \ref{lemma:4}, \ref{lemma:err8},  \ref{prop:err2},  \ref{prop:err31} and \ref{prop:err}.

\subsection*{Errors in the stable interval}

We denote by $\Pe_{T_k,x_k,p_k}$ the law of $(x(t),p(t))$ given
$x(T_k)=x_k$, $p(T_k)=p_k$. In this first part of the Section we will
prove the following Proposition.

\begin{prop}\label{prop:y,R}
There exists $C>0$ such that, for any $\zeta>0$, $\xi,\eps$
small enough,
\begin{equation}\label{err5}
\mathbf 1_{(x_k,p_k) \in \mathcal
H_k^\xi}\;\Pe_{T_k,x_k,p_k}\;\left\{\sup_{T_k\le t \le S_k}|
y_k(t)|\ge \frac{\eps} {\eta_\eps}\: \eps^{-\zeta}\right\}\le
e^{-C \eps^{-2\zeta}}
\end{equation}
\begin{equation}\label{err8}
\mathbf 1_{(x_k,p_k) \in \mathcal
H_k^\xi}\;\Pe_{T_k,x_k,p_k}\;\left\{\sup_{T_k\le t \le S_k}|\mathcal
R_k(t)| \ge \frac {\eps \sigma_\eps} {\eta_\eps^{2}} \:
\eps^{-2\zeta}\right\}\le e^{-C \eps^{-2\zeta}}
\end{equation}
and
\begin{equation}\label{err8'}
\mathbf 1_{(x_k,p_k) \in \mathcal
H_k^\xi}\;\Pe_{T_k,x_k,p_k}\;\left\{\sup_{T_k\le t \le S_k}|r_k(t)|
\ge \sigma_\eps \;\eps^{-\zeta}\right\}\le e^{-C \eps^{-2\zeta}}
\end{equation}

\end{prop}

\vskip1cm

\begin{lemma}\label{lemma:err1}
There exists $C>0$ such that, for any $\zeta>0$,  $\xi,\eps$ small
enough,
\begin{equation}\label{err5'}
\mathbf 1_{(x_k,p_k) \in \mathcal
H_k^\xi}\;\Pe_{T_k,x_k,p_k}\left\{\sup_{T_k\le t \le \bar S_k}|
y_k(t)|\ge\frac{\eps}{\eta_\eps}\, \eps^{-\zeta} \right\}\le
e^{-C\eps^{-2\zeta}}
\end{equation}
\end{lemma}

\vskip.5cm

{\bf Proof.}  Suppose that $(x_k,p_k) \in \mathcal H_k^\xi$ for some
$\xi>0$ small enough. Let us fix $\zeta>0$, then, from Lemma \ref{lemma:y},
\eqref{MS} and \eqref{err1} it follows that there exists $C>0$
such that, for any $\eps$ small enough,
\begin{equation}\label{err2}
\Pe_{T_k,x_k,p_k}\left\{\sup_{T_k\le t \le \bar S_k}|\bar y_k(t)|\ge
\frac{\eps}{2\eta_\eps}\, \eps^{-\zeta}\right\}\le e^{-C\eps^{-2\zeta}}
\end{equation}

We recall that $y_k(t)= \bar y_k(t)+ \mathcal Y_k(t)$ with 
\begin{equation}\label{err3'}
\mathcal Y_k(t) \le H_y(T_k,t)  \cdot \sup_{T_k \le s \le \bar{S_k}} \varphi(X_k(s),y_k(s)) \quad \quad \text{for} \quad \quad T_k \le t \le \bar S_k
\end{equation}
(see \eqref{Y} and \eqref{Hy}).
From \eqref{phi},
 we know that $\varphi(x,y)=\mathcal O(y^2)$ for small $|y|$, then if we define  the
stopping time $\Upsilon_k^{\zeta}:=\inf\{t \ge T_k: |y_k(t)|\ge
\eps^{1-\zeta} \eta^{-1}_\eps\}$,
 then there exists $c>0$ such
that
\begin{equation}\label{err3}
\sup_{T_k \le s \le \Upsilon_k^{\zeta} \wedge \bar S_k}|\varphi(X_k(s),y_k(s))|\le c  \, \frac
{\eps^{2(1-\zeta)}}{\eta_\eps^{2}}
\end{equation}
thus, from \eqref{err3'} and \eqref{Hy} we have
\begin{equation}\label{err4}
|\mathcal Y_k(t)| \le c  \, \frac
{\eps^{2(1-\zeta)}}{\eta_\eps^3}
\quad \text{for any} \quad T_k \le t \le \Upsilon_k^{\zeta}\wedge \bar S_k.
\end{equation}

Thus, since $|y_k(t)|\le |\bar y_k(t)|+ |\mathcal Y_k(t)|$, by
\eqref{err2} and \eqref{err4}, 
 with $\Pe_{T_k,x_k,p_k}$-probability greater than $1-2
 e^{-C \eps^{-2\zeta}}$,
\begin{equation*}
\sup_{T_k \le t \le \Upsilon^\zeta_k \wedge \bar S_k}|y_k(t)|\le
\frac{\eps^{1-\zeta}}{2\eta_\eps} + c \, \frac{\eps^{2(1-\zeta)}}
{\eta_\eps^{3}} < \frac{\eps^{1-\zeta}}{\eta_\eps},
\end{equation*}
hence, in particular, with the same probability, $\bar S_k <
\Upsilon^\zeta_k$, then \eqref{err5'} follows. \qed

\vskip1cm

\begin{lemma}\label{lemma:err2}
There exists $C>0$ such that, for any  $\zeta>0$, $\xi,\eps$ small
enough,
\begin{equation}\label{err8}
\mathbf 1_{(x_k,p_k) \in \mathcal H_k^\xi}\;\Pe_{T_k,x_k,p_k}\left\{
\sup_{T_k\le t \le \bar S_k}|\mathcal R_k(t)| \ge  \frac{\eps\sigma_\eps}
{\eta_\eps^{2}}\, \eps^{-2\zeta}\right\}\le e^{-C \eps^{-2\zeta}}
\end{equation}
\end{lemma}

\vskip.5cm

{\bf Proof.} Suppose that $(x_k,p_k) \in \mathcal H_k^\xi$ for some
$\xi$ small enough. From \eqref{R} we know that 
\begin{equation}\label{err6'}
\mathcal R_k(t)= H_r(T_k,t)\cdot \sup_{T_k \le t \le \bar S_k}\varphi(X_k(s),y_k(s))\, ds
\end{equation}
thus, by Lemma
\ref{lemma:err1}, we know that there exists $c>0$ such that
\begin{equation}\label{err3''}
\sup_{T_k \le s \le \bar S_k}|\varphi(X_k(s),y_k(s))|\le c  \, \frac
{\eps^{2(1-\zeta)}}{\eta_\eps^{2}}
\end{equation}
with $\Pe_{T_k,x_k,p_k}$-probability greater than $1-e^{-C
\eps^{-2\zeta}}$, then, from \eqref{err6'}, \eqref{err3''} and \eqref{Hr} it follows that
\begin{equation}\label{err6}
 \sup_{T_k \le t \le \bar S_k}|\mathcal R_k(t)|\le c\;
\frac{\eps^{2(1-\zeta)}}{\eta_\eps^2} \;\eta_\eps^{-\frac{1}{1+\theta}}
\end{equation}
with $\Pe_{T_k,x_k,p_k}$-probability greater than $1-e^{-C
\eps^{-2\zeta}}$, hence \eqref{err8} follows from the definition of $\sigma_\eps$. \qed

\vskip1cm

\begin{lemma}\label{lemma:err3}
There exists $C>0$ such that, for any $\zeta>0$, $\xi,\eps$ small
enough,
\begin{equation}\label{err11}
\mathbf 1_{(x_k,p_k) \in \mathcal
H_k^\xi}\;\Pe_{T_k,x_k,p_k}\left\{\sup_{T_k\le t \le \bar S_k}|r_k(t)|\ge \sigma_\eps  \, \eps^{-\zeta}\right\}\le
e^{-C\eps^{-2\zeta}}
\end{equation}
\end{lemma}

\vskip.5cm

{\bf Proof.} Assume $(x_k,p_k) \in \mathcal H_k^\xi$,  with $\xi$ small enough. We recall that $r_k(t)= \bar r_k(t)+ \mathcal R_k(t)$. From Lemma
\ref{lemma:r}, \eqref{MS} and \eqref{err9}, there exists $C>0$
such that, for any $\zeta>0$, $\eps$ small enough,
\begin{equation}\label{err10}
\Pe_{T_k,x_k,p_k}\left\{\sup_{T_k\le t \le \bar S_k}|\bar r_k(t)|\ge
 \frac{\sigma_\eps} 2 \; \eps^{-\zeta}\right\}\le  e^{-C\eps^{-2\zeta}}
\end{equation}
hence the result descends from \eqref{err10} and Lemma
\ref{lemma:err2}, since, from \eqref{cond2}, $\eps =o(\eta_\eps^{2})$.  \qed

\vskip1cm

\begin{coro}\label{coro:err1}
There exists $C>0$ such that, for any  $\zeta>0$, $\xi,\eps$ small
enough,
\begin{equation}\label{err12}
\mathbf 1_{(x_k,p_k) \in \mathcal
H_k^\xi}\;\Pe_{T_k,x_k,p_k}\left\{\sup_{T_k\le t \le \bar S_k}|
q_k(t)|\ge \frac{\eps}{\eta_\eps}\, \eps^{-\zeta}\right\}\le e^{-C
\eps^{-2\zeta}}
\end{equation}
\end{coro}

\vskip.5cm

{\bf Proof.} We recall that $q_k(t)= r_k(t)+\o_k(t)
y_k(t)$, then \eqref{err12}  easily follows from Lemma \ref{lemma:err1}, Lemma
\ref{lemma:err3} and Lemma \ref{omega}. \qed

\vskip1cm

\begin{lemma}\label{lemma:err6}
Let $\bar S_k$ be the stopping time defined in \eqref{S} and
\begin{equation}\label{S'}
\tilde S_k:=\inf\left\{t \ge T_k: X_k(t)\ge 2k \pi -
\frac{2\eta_\eps}{(\lap-\lam)}\right\}
\end{equation}

then there exists $C>0$ such that, for any  $\zeta>0$, $\xi,\eps$ small enough,
\begin{equation}\label{H1}
\mathbf 1_{(x_k,p_k) \in \mathcal H_k^\xi}\;\Pe_{T_k,x_k,p_k}\left\{\tilde S_k \le S_k \le \bar S_k\right\}\ge 1-e^{-C \eps^{-2\zeta}}
\end{equation}
\end{lemma}

\vskip.5cm

{\bf Proof.} 
We only show that
\begin{equation}\label{SS}
\mathbf 1_{(x_k,p_k) \in \mathcal H_k^\xi}\;\Pe_{T_k,x_k,p_k}\left\{S_k
\le \bar S_k\right\}\ge 1-e^{-C \eps^{-2\zeta}}
\end{equation}
since the arguments for the estimate of
$\Pe_{T_k,x_k,p_k}\left\{\tilde S_k \le S_k \right\}$ are specular.
 Assume $(x_k,p_k) \in \mathcal
H_k^\xi$, for some $\xi$ small enough. We have 
\begin{equation}\label{err130}
\Pe_{T_k,x_k,p_k}\left\{S_k \le
\bar S_k\right\} \ge \Pe_{T_k,x_k,p_k}\left\{v_k(\bar S_k)\le
\eta_\eps\right\}
\end{equation}
 where
\begin{equation}\label{err13'}
v_k(t)= P_k(t)-\lap(X_k(t)-2k\pi) + q_k(t)-\lap y_k(t).
\end{equation}
Recalling  the definition of $\bar S_k$ and $x_k'$ in \eqref{S},  we have
\begin{equation}\label{err13''}
v_k(\bar S_k)= \wp_k(x_k')-\lap(x_k'-2k\pi) + q_k(\bar S_k)-\lap y_k(\bar S_k).
\end{equation}
thus, from Lemma \ref{lemma:cond'} we have
\begin{equation}\label{err13}
\Big|v_k(\bar S_k)-\frac{\eta_\eps}2 -(q_k(\bar S_k)-\lap
y_k(\bar S_k))\Big| \le c \, \tilde  \sigma_\eps \eps^{-\xi}
\end{equation}
for some $c>0$. By Lemma \ref{lemma:err1} and Corollary
\ref{coro:err1}, there exists $C>0$ such that, for any $\zeta,\eps$ small enough,
\begin{equation}\label{err14}
\Pe_{T_k,x_k,p_k}\left\{ |q_k(\bar S_k)-\lap y_k(\bar S_k)|\ge
\frac{\eps}{\eta_\eps}\, \eps^{-\zeta}\right\}\le \eps^{-C
\eps^{-2\zeta}}
\end{equation}
 From \eqref{cond2} we know that $\tilde \sigma_\eps \eps^{-\xi} =o(\eta_\eps)$ and $\eps^{1-\xi}
 =o(\eta^2_\eps)$, for $\xi$ small enough, thus,
 from \eqref{err13} and
\eqref{err14} it follows that
\begin{equation}\label{err14'}
\Pe_{T_k,x_k,p_k}\left\{ v_k(\bar S_k)> \eta_\eps\right\}\le \eps^{-C
\eps^{-2\zeta}}
\end{equation}
then \eqref{SS} follows from  \eqref{err130} and \eqref{err14'}. \qed

\vskip1cm

{\bf Proof of Proposition \ref{prop:y,R}.} It directly follows
from \eqref{err5'}, \eqref{err8}, \eqref{err11} and Lemma
\ref{lemma:err6}. \qed

\subsection*{Errors in the critical interval}

For any $\xi>0$ we recall that
\begin{equation}
\mathcal K_{k}^{\xi}=\left\{ (x,p)\;: \:
p-\lap(x-2k\pi)=\eta_\eps, \: |p-\lam(x-2k\pi)|\le \sigma_\eps
\, \eps^{-\xi}\right\}.
\end{equation}
and denote by $\Pe_{S_k,\bar x_k,\bar p_k}$ the law of $(x(t), p(t))$ given
$x(S_k)=\bar x_k$, $p(S_k)=\bar p_k$. We will prove the following result.

\vskip1cm

\begin{prop}\label{prop:err1}
 We have
\begin{equation}\label{err33}
\sup_{S_k \le t \le T_{k+1}}|z_k(t)|\le \eta_\eps \quad \quad
\Pe_{S_k,\bar x_k,\bar p_k} \:\text{a.s.}
\end{equation}
moreover there exists $C>0$ such that, for any $\d>0$,
$\zeta,\eps$ small enough,
\begin{equation}\label{err33'}
\Pe_{S_k,\bar x_k,\bar p_k}\left\{ \sup_{S_k \le t \le T_{k+1}}|v_k(t)|\ge
 (1+\d) \eta_\eps\right\}\le e^{-C
\eps^{-2\zeta}}
\end{equation}
for any $(\bar x_k,\bar p_k)\in \mathbb{R}^2$.
\end{prop}

\vskip1cm

\begin{lemma}\label{lemma:err4}
There exists $C>0$ such that, for any $\d,\zeta>0$, $\eps$ small
enough,
\begin{equation}\label{err21}
\Pe_{S_k,\bar x_k,\bar p_k}\;\left\{\sup_{t \ge S_k}|\bar v_k(t)|\le
(1+\d)\eta_\eps\right\}\ge 1-e^{-C \eps^{-2 \zeta}}
\end{equation}
for any  $(\bar x_k,\bar p_k)\in \mathbb{R}^2$.

\end{lemma}

\vskip.5cm

{\bf Proof.}  From Lemma \ref{lemma:v} we know that the
probability law of $\bar v_k(S_k+t)-\eta_\eps e^{\lam t}$ is
 a centered Gaussian  of variance
$\sigma_v^2(t)$, independently on the initial condition $(\bar x_k,\bar p_k)$ at time $S_k$. 
By \eqref{sigmav} we have
\begin{equation}\label{sigmav'}
 \sup_{t \ge 0}\sigma_v^2(t)\le \frac
{\eps^2} {2 |\lam|}
\end{equation}
thus, by \eqref{MS}, there exists $C>0$ such that,
\begin{equation}\label{err17}
\Pe_{S_k,\bar x_k,\bar p_k}\left\{\sup_{t \ge 0}\big|\bar v_k(S_k+t)-\eta_\eps
\,e^{\lam t}\big|\ge \eps^{1-\zeta}\right\}\le
e^{-C\eps^{-2\zeta}}
\end{equation}
for any $\zeta,\d>0$, $\eps$ small enough, $(\bar x_k,\bar p_k)\in
\mathbb{R}^2$, then \eqref{err21} follows since, from
\eqref{cond2}, $\eps^{1-\zeta}=o(\eta_\eps)$. \qed

\vskip1cm

\begin{lemma}\label{lemma:err5}
There exists $C>0$ such that, for any $\xi>0$, $\zeta,\eps$ small
enough,
\begin{equation}\label{err23}
\mathbf 1_{(\bar x_k,\bar p_k)\in \mathcal K_k^\xi}\;\Pe_{S_k,\bar x_k,\bar p_k}\left\{
\sup_{t \ge S_k}|\bar z_k(t)|\, e^{-\lap(t-S_k)}\ge \sigma_\eps\,
\eps^{-\xi}\right\}\le e^{-C \eps^{-2\zeta}}
\end{equation}
\end{lemma}

\vskip.5cm

{\bf Proof.} From Lemma \ref{lemma:z''} we know that the
probability law of the process $\bar z_k(S_k+t)-z_k(S_k)e^{\lap
t}$  is a centered Gaussian of variance $\sigma_z^2(t)$
independent of the values $(\bar x_k,\bar p_k)$ at time $S_k$. From the formula
in \eqref{sigmaz} we have
\begin{equation}\label{sigmaz,v}
\sup_{t\ge 0}\sigma^2_z(t) \,e^{-2\lap t} \le \frac {\eps^2} {2
\lap}
\end{equation}
thus, by \eqref{MS}, there exists $C>0$ such that, for any
$\zeta>0$, $\eps$ small enough, $(\bar x_k,\bar p_k)\in \mathbb{R}^2$,
\begin{equation}\label{err18}
\Pe_{S_k,\bar x_k,\bar p_k}\left\{\sup_{t \ge 0}\big|\bar z_k(S_k+t)\,e^{-\lap
t}-z_k(S_k)\big|\ge  \eps^{1-\zeta}\right\}\le e^{-C
\eps^{-2\zeta}}
\end{equation}

then, with $\Pe_{S_k,\bar x_k,\bar p_k}$-probability larger than $1-e^{-C
\eps^{-2\zeta}}$
\begin{equation}\label{err18'}
 |\bar z_k(t)|e^{-\lap (t-S_k)}\le
|z_k(S_k)|+ \eps^{1-\zeta}\quad \quad \forall \,t \ge S_k.
\end{equation}
Let us assume $(\bar x_k,\bar p_k)\in \mathcal K_k^\xi$, for some $\xi>0$, hence
\eqref{err23} follows since $|z_k(S_k)|=|p-\lam(x-2k\pi)|\le
\sigma_\eps \, \eps^{-\xi}$ and since, from
\eqref{cond2}, $\eps^{1-\zeta}=o(\sigma_\eps\, \eps^{-\xi})$, for
$\zeta$ small enough. \qed

\vskip2cm

{\bf Proof of Propositions \ref{prop:err1}, \ref{prop:err2} and
\ref{prop:err31}.} \eqref{err33} follows directly  from the
definitions of $T_{k+1}$ and $\mathcal K_k^\xi$, since
$\sigma_\eps \,
\eps^{-\xi} =o(\eta_\eps)$ for $\xi$ small enough.\\
Assume $(\bar x_k,\bar p_k)\in \mathcal K_k^\xi$, thus $|x-2k\pi|< 3
\eta_\eps/(\lap-\lam)$. Consider the stopping time
\begin{equation*}
\Gamma_k:=\inf \left\{t \ge S_k: |x(t)-2k\pi|\ge \frac{3
\eta_\eps}{\lap-\lam}\right\},
\end{equation*}
then, since $\psi_k(x)=\mathcal O((x-2k\pi)^2)$ for small
$(x-2k\pi)$, by \eqref{V}  there exists $c>0$ such
that, for any $S_k \le t \le \Gamma_k \wedge T_{k+1}$,
\begin{equation}\label{err26}
|\mathcal V_k(t)|\le c \,\eta_\eps^2 \,e^{\lam t}\int_{S_k}^t
e^{-\lam s}\; ds \le c \,\eta_\eps^2
\end{equation}
and
\begin{equation}\label{err25}
|\mathcal Z_k(t)|\le c \,\eta_\eps^2\, e^{\lap t}\int_{S_k}^t
e^{-\lap s}\; ds \le c \,\eta_\eps^2 \,e^{\lap (t-S_k)}
\end{equation}

then
\begin{equation}\label{err27}
\sup_{S_k \le t \le \Gamma_k\wedge T_{k+1}}|\mathcal V_k(t)| \le
c' \eta_\eps^2 \quad \text{and}\quad \sup_{S_k \le t \le \Gamma_k
\wedge T_{k+1}}|\mathcal Z_k(t)| e^{-\lap (t-S_k)}\le c'
\eta_\eps^2
\end{equation}
for suitable $c'>0$. Recalling that $z_k(t)=\bar z_k(t)+\mathcal
Z_k(t)$, since $\eta_\eps^2=o(\sigma_\eps)$, it follows from
\eqref{err23} and the right hand side of \eqref{err27} that, for
any $\xi>0$, $\zeta,\eps$ small enough,
\begin{equation}\label{err30}
\mathbf 1_{(\bar x_k,\bar p_k)\in \mathcal
K_k^\xi}\;\Pe_{S_k,\bar x_k,\bar p_k}\left\{\sup_{S_k \le t \le \Gamma_k \wedge
T_{k+1}} |z_k(t)|\,e^{-\lap (t-S_k)}\ge \sigma_\eps \,
\eps^{-\xi} \right\}\le e^{-C\eps^{-2\zeta}}
\end{equation}
for a suitable $C>0$.

On the other hand we have $v_k(t)=\bar v_k(t)+\mathcal V_k(t)$,
then, by Lemma \ref{lemma:err4} and \eqref{err27}, there exists
$C>0$ such that, for any $(\bar x_k,\bar p_k)\in \mathbb{R}^2$, $\zeta,\d>0$,
$\eps$ small enough,
\begin{equation}\label{err29}
\Pe_{S_k,\bar x_k,\bar p_k}\left\{\sup_{S_k \le t \le \Gamma_k \wedge
T_{k+1}}|v_k(t)|\ge \eta_\eps(1+\d)\right\}\le e^{-C
\eps^{-2\zeta}}
\end{equation}

We recall that $x(t)-2k\pi= (z_k(t)-v_k(t))/(\lap-\lam)$, thus from \eqref{err33} and \eqref{err29} it follows that
\begin{equation}\label{err30'}
\Pe_{S_k,\bar x_k,\bar p_k}\left\{\Gamma_k> T_{k+1} \right\}\ge
\Pe_{S_k,\bar x_k,\bar p_k}\left\{\sup_{S_k \le t \le \Gamma_k \wedge T_{k+1}}|x(t)-2k\pi|\le
\frac{(2+\d) \eta_\eps}{\lap-\lam}\right\} \ge 1-e^{-C
\eps^{-2\zeta}}
\end{equation}
thus \eqref{err33'} follows from \eqref{err29} and \eqref{err30'},
whereas \eqref{err31} follows from \eqref{err30} and
\eqref{err30'}.
 \eqref{err32'} and \eqref{err32} descend both from
\eqref{err26} and \eqref{err30'}. \qed

\subsection*{Conclusion of the Proofs}

We conclude the proofs of Propositions \ref{lemma:err8}, \ref{lemma:4} and \ref{prop:err}.

\vskip1cm

\begin{prop}\label{prop:compl}
Suppose that $(x_k,p_k) \in \mathcal H_k^\xi$ with $\xi$ small enough, then there exist $c'>c>0$ such that 
\begin{equation}\label{f*''''}
c \,\frac{\eta_\eps^{(1+\theta)}}{\eps}\le \frac{|P_k(S_k)-\wp^*_k(X_k(S_k))|}{\sigma_r(T_k,S_k)|\wp_k(\bar x)-\wp_k^*(\bar x)|} \le c' \frac{\eta_\eps^{(1+\theta)}}{\eps}
\end{equation}
for any $\eps$ small enough.
\end{prop}
\vskip.5cm
{\bf Proof.} Suppose that $(x_k,p_k)\in \mathcal H_k^\xi$. From \eqref{SSS}
we know that
\begin{eqnarray}\label{A3}
\sigma_r^2(T_k,S_k)= \eps^2 \Sigma_r^{(2)}[\wp_k,\bar x](X_k(S_k))
\end{eqnarray}
where, by \eqref{f*'},
\begin{equation}\label{f*''}
\lim_{\eps\to
0}\frac{\Sigma_r^{(2)}[\wp_k,\bar x](X_k(S_k))}{[P_k(S_k)-\wp^*_k(X_k(S_k))]^2}\;[\wp_k(\bar x)-\wp_k^*(\bar x)]^2\,\(\bar x-2(k-1)\pi\)^{2(1+\theta)}\in
(0,+\i)
\end{equation}

then, from \eqref{x01'} it follows that 
\begin{equation}\label{f*'''}
\lim_{\eps \to 0}\frac{|P_k(S_k)-\wp^*_k(X_k(S_k))|}{\sigma_r(T_k,S_k)|\wp_k(\bar x)-\wp_k^*(\bar x)|} \; \frac{\eps}{\eta_\eps^{(1+\theta)}} \in (0,+\i)
\end{equation}

from which the result. \qed

\vskip1cm

\begin{lemma}\label{lemma:H}
There exist $C,c,c'>0$ such that, for any $\zeta>0$, $\xi,\eps$
small enough,
\begin{equation}\label{H''}
\mathbf 1_{(x_k,p_k)\in \mathcal H_k^\xi}\;\Pe_{T_k,x_k,p_k}\big\{c
\,\eta_\eps<2k\pi-X_k(S_k)\le c'\eta_\eps\big\}\ge 1 -e^{-C
\eps^{-2\zeta}}
\end{equation}
\end{lemma}

\vskip.5cm {\bf Proof.} \eqref{H''} follows from  Lemma
\ref{lemma:err6} and the monotonicity of $X_k(t)$. \qed

\vskip1cm

{\bf Proof of Proposition \ref{lemma:4}.} 
From Lemma \ref{lemma:H} the hypothesis \eqref{x01} holds for the couple $\bar x,X_k(S_k)$ with probability larger than $1 -e^{-C
\eps^{-2\zeta}}$ then  \eqref{H3} directly follows from \eqref{A3} and \eqref{g**}. \qed

 \vskip1cm

{\bf Proof of Proposition \ref{lemma:err8}.}
\eqref{A} easily follows from \eqref{f*''''} and \eqref{p0''} since, from \eqref{sigma} and \eqref{bsigma} we have  $\eps^{-1}\eta_\eps^{1+\theta} \bar \sigma_\eps = \eps^{\theta}$.  \eqref{A1} is a direct consequence of \eqref{A} and \eqref{H3}\qed

\vskip1cm

Let us define the processes
\begin{equation*}
Z_k(t)=\wp_k^*(X_k(t))-\lam(X_k(t)-2k\pi), \quad
V_k(t)=\wp_k^*(X_{k}(t))-\lap(X_{k}(t)-2(k-1)\pi)
\end{equation*}
we have the following result.

\vskip1cm

\begin{lemma}
Let $(x_k,p_k)\in \mathcal H_{k}^\xi$ for some $\xi>0$, then there
exists $c>0$ such that
\begin{equation}\label{err34}
 |V_{k}(T_{k})|\le c \,\eta_\eps^2\quad \quad \Pe_{T_{k},x_k,p_k}\: \text{a.s.}
\end{equation}
There exist $C,c'>0$ such that, for any $\xi>0$, $\zeta,\eps$
small enough,
\begin{equation}\label{err34'}
\mathbf 1_{(x_k,p_k)\in \mathcal
H_k^\xi}\;\Pe_{T_k,x_k,p_k}\left\{|Z_k(S_k)|\le c'
\eta_\eps^2\right\}\ge 1-e^{-C \eps^{-2\zeta}}
\end{equation}

\end{lemma}

\vskip.5cm

 {\bf Proof.}
Recalling that $X_k(T_k)=x(T_k)=\bar x$, \eqref{err34} follows from
\eqref{p'*+} and the left hand side of \eqref{x01'}, whereas \eqref{err34'} follows from
\eqref{p'*-} and \eqref{H''}. \qed

\vskip1cm

\begin{lemma}\label{lemma:err7}
There exist $C,c>0$ such that, for any $\zeta>0$, $\xi,\eps$ small
enough,
\begin{equation}\label{err51}
\mathbf 1_{(x_k,p_k)\in \mathcal
H_k^\xi}\;\Pe_{T_k,x_k,p_k}\left\{|\o_k(S_k)-\lam|\le  c \,
\frac{\bar \sigma_\eps}{\eta_\eps}\right\} \ge 1- e^{-C
\eps^{-2\zeta}}
\end{equation}
\end{lemma}

\vskip.5cm

{\bf Proof.}
We recall that $\o_k(S_k)=\frac d {dx}\,
\wp_k(X_k(S_k))$. Suppose that $(x_k,p_k)\in \mathcal H_k^\xi$, $\xi$ small enough then, by Lemma \ref{lemma:H} the hypothesis \eqref{x01} holds for the couple $\bar x,X_k(S_k)$ with probability larger than $1 -e^{-C
\eps^{-2\zeta}}$. Hence we can apply  \eqref{F10} and \eqref{p0''} and obtain that
 there exist $c,C>0$ such that  for any $\zeta,\eps$ small enough,
\begin{equation}\label{err52}
\mathbf 1_{(x_k,p_k)\in \mathcal
H_k^\xi}\;\Pe_{T_k,x_k,p_k}\left\{\Big|\o_k(S_k)-\frac d {dx}\,\wp_k^*(X_k(S_k))\Big|\le c \,
\frac{\bar \sigma_\eps} {\eta_\eps} \, \eps^{-\xi}\right\} \ge 1- e^{-C
\eps^{-2\zeta}}
\end{equation}
on the other hand, by \eqref{p'*-} ,  there exist
$c',C>0$ such that, for any $\zeta,\eps$ small enough,
\begin{equation}\label{err53}
\mathbf 1_{(x_k,p_k)\in \mathcal
H_k^\xi}\;\Pe_{T_k,x_k,p_k}\left\{\Big|\frac d
{dx}\,\wp_k^*(X_k(S_k))-\lam\Big|\le c'
 \,\eta_\eps\right\}\ge 1- e^{-C
\eps^{-2\zeta}}
\end{equation}
then \eqref{err51} follows from \eqref{err52} and \eqref{err53} since
$\eta_\eps^2=o(\bar \sigma_\eps)$. \qed

\vskip1cm

{\bf Proof of Proposition \ref{prop:err}.} Suppose that $(x_k,p_k)\in \mathcal H_k^\xi$ for some $\xi>0$, then, 
from the definitions of $z_k(t)$ in \eqref{z,v}, $\hat z_k(t)$ in \eqref{hatz} and $r_k(t)$ in \eqref{rr} we have
\begin{eqnarray}
z_k(S_k)-\hat z_k(S_k)&=&Z_k(S_k)+\lam q_k(S_k)-\omega_k(S_k) y_k(S_k)\\
&=&Z_k(S_k)+r_k(S_k)+(\omega_k(S_k)-\lam)y_k(S_k)\\
&=&\mathcal R_k(S_k)+Z_k(S_k) +(\o_k(S_k)-\lam)y_k(S_k)
\end{eqnarray}
where the last equivalence follows since $\bar r_k(S_k)=0$.
Hence \eqref{properr-} descends from \eqref{err5}, \eqref{err8},
\eqref{err34'} and Lemma \ref{lemma:err7}.\\ Suppose now that $(\bar x_{k-1},\bar p_{k-1})\in \mathcal K_{k-1}^\xi$ for some $\xi>0$. 
From the definitions of $v_k(t)$ in \eqref{z,v} and $\mathcal V_k(t)$ in \eqref{V}
we have
\begin{eqnarray*}
p(T_k)-\wp_k^*(x(T_k))-\bar v_{k-1}(T_k)&=&\mathcal
V_{k-1}(T_k)- V_{k}(T_k)+\lap y_k(T_k)+\wp^*_k(X_k(T_k))-\wp_k^*(x(T_k))\\
&=&\mathcal V_{k-1}(T_k)- V_{k}(T_k)
\end{eqnarray*}
where the last equality follows since $X_k(T_k)=x(T_k)=x_k$. Then \eqref{properr+} descends from \eqref{err32} and
\eqref{err34}. \qed

\section{Appendix}

In the present Appendix we
provide  a Gaussian Inequality and a comparison result.

\vskip.5cm

\paragraph{Marcus-Shepp inequality for Gaussian processes.}
There is a classical result of Landau and Shepp \cite{LS} and
Marcus and Shepp \cite{MS} that gives an estimate for the supremum of a general centered Gaussian
process.
If
$X(t)$ is an a.s. bounded, centered Gaussian process of variance
$\sigma^2(t)$, then,
\begin{equation}\label{MS'}
\lim_{\la \to \i}\frac 1 {\la^{2}}\ln \P \left\{\sup_{t \in I}
X(t) \ge \la\right\}=-\frac 1 {2 \sigma^2_I} \quad \text{with}
\quad \sigma^2_I:=\sup_{t \in I}\sigma^2(t)
\end{equation}
An immediate consequence of \eqref{MS'} is that for any $\la$
large enough, $\d$ small enough,
\begin{equation}\label{MS''}
\P \left\{\sup_{t\in I} X(t) \ge \la\right\}\le  e^{-\frac
{\la^2}{2\sigma^2_I}(1-\d)}
\end{equation}
moreover, by symmetry we have
\begin{equation}\label{MS'''}
\P \left\{\sup_{t\in I} |X(t)| \ge \la\right\}\le  2\P
\left\{\sup_{t\in I} X(t) \ge \la\right\}.
\end{equation}
By applying the result to the process $X(t)/\sigma(t)$ we get
\begin{equation}\label{MS}
\P \left\{\sup_t \frac {|X(t)|}{\sigma(t)} \ge \la\right\}\le 2
e^{-\frac {\la^2}2(1-\d)}
\end{equation}
for $\la$ large enough, $\d$ small enough.

\vskip.5cm

\paragraph{Comparison with Gaussian Processes.}

In our proofs we repeatedly make use of a comparison argument
comparing the solution of a linear SDE with the solution of a more
general SDE, let us see.

Let $X_t$ be a solution of the problem
\begin{equation}\label{XXX}
 dX_t=(a(t)X_t+ b(t))dt+\xi dw_t,
\end{equation}
with $ a, b: \mathbb{R}^+ \ra \mathbb{R}$ bounded on bounded
intervals and $\xi \in \mathbb{R}$, then $X(t)$ is a Gaussian
process of the form
\begin{equation*}
X(t)=X(t_0)\,e^{\int_{t_0}^t a(s)\, ds}+\int_{t_0}^t b(s)\,
e^{\int_s^t a(s')\, ds'}\, ds + \xi \int_{t_0}^t e^{\int_s^t
a(s')\,ds'}dw_s.
\end{equation*}

Consider, now, the processes $x(t)$ solution of
\begin{equation*}
 dx_t= c(x_t,t)dt +\xi dw_t
\end{equation*}
with the same noise of \eqref{XXX},   $c: \mathbb{R} \times
\mathbb{R}^+ \ra \mathbb{R}$ globally Lipschitz.

\vskip1cm

\begin{lemma}\label{lemma:comp}
For $X(t),x(t)$ as above we define
$\d_t:=c(X_t,t)-[a(t)X_t+b(t)]$, $\Delta_t:=X_t-x_t$, and let
$\tau \in \mathbb{R}^+$ be a generic random variable. Suppose
\begin{equation*}
 \sgn (\Delta_\tau) = \sgn
(\d_\tau) \quad \text{ or }\quad \Delta_\tau=0,
\end{equation*}
then
\begin{equation*}
\sgn (\Delta_t) = \sgn (\d_t) \quad \text{for any} \quad  \tau \le
t \le \inf\{s \ge \tau: \d_s=0\} \quad \text{a.s.}
\end{equation*}
\end{lemma}

\vskip.5cm

{\bf Proof.} We have
\begin{equation*}
d\Delta_t=(a(t)\Delta_t+\d_t)dt
\end{equation*}
thus, for any $\tau \ge 0$
\begin{equation*}
\Delta(t)=\Delta(\tau)\, e^{\int_\tau^t  a(s)\,
ds}+\int_\tau^t\d(s)\, e^{\int_s^t a(s')\, ds'}\, ds
\end{equation*}
then follows the result. \qed

\end{document}